\documentclass[english,12pt,oneside]{article}
\pdfoutput=1
\usepackage[T1]{fontenc}
\usepackage[latin9]{inputenc}
\usepackage[usenames,dvipsnames]{xcolor}
\usepackage{color}
\usepackage{babel}
\usepackage{setspace}
\usepackage{amstext}
\usepackage{amssymb}
\PassOptionsToPackage{normalem}{ulem}
\usepackage{ulem}
\usepackage[unicode=true,pdfusetitle,
 bookmarks=true,bookmarksnumbered=false,bookmarksopen=false,citecolor=Turquoise,
 breaklinks=false,pdfborder={0 0 1},backref=false,colorlinks=true,pdfpagemode=FullScreen,linktoc=page]
 {hyperref}
\usepackage{graphicx}
\usepackage{mathtools}
\usepackage[footnotesize,sc,format=plain,indention=.00075cm,width=.9\textwidth]{caption}
\definecolor{note_fontcolor}{rgb}{0.80078125, 0.80078125, 0.80078125}
\usepackage[top=3 cm, bottom=3 cm, left=3.1416 cm, right=3.1416 cm]{geometry}

\def\beq{\begin{equation}}
\def\eeq{\end{equation}}
\def\bea{\begin{eqnarray}}
\def\eea{\end{eqnarray}}
\def\beq{\begin{equation}}
\def\eeq{\end{equation}}
\def\bea{\begin{eqnarray}}
\def\eea{\end{eqnarray}}
\def\bit{\begin{itemize}}
\def\eit{\end{itemize}}

\def\misse{E\hspace{-0.25cm}/_{T}}
\def\met{E\hspace{-0.25cm}/_{T}}
\newcommand{\prd}{Phys.Rev.D}
\newcommand{\Eapeak}{E_{a,{\textnormal{{\footnotesize peak}}}}}
\newcommand{\Ebpeak}{E_{b,{\textnormal{{\footnotesize peak}}}}}

\renewcommand{\url}[1]{\href{#1}{\textsc{\textcolor{Emerald}{WebSite}}}}
\newcommand{\tg}{\tilde{g}}
\newcommand{\tb}{\tilde{b}}
\newcommand{\gev}{\textrm{ GeV}}
\newcommand{\GeV}{\textrm{ GeV}}

\newcommand{\TeV}{\textrm{ TeV}}

\newcommand{\eq}[1]{Eq.~(\ref{#1})}
\newcommand{\eqs}[1]{Eqs.~(\ref{#1})}

\long\def\symbolfootnote[#1]#2{\begingroup%
\def\thefootnote{\fnsymbol{footnote}}\footnote[#1]{#2}\endgroup}

\newenvironment{changemargin}[2]{%
\begin{list}{}{%
\setlength{\topsep}{0pt}%
\setlength{\leftmargin}{#1}%
\setlength{\rightmargin}{#2}%
\setlength{\listparindent}{\parindent}%
\setlength{\itemindent}{\parindent}%
\setlength{\parsep}{\parskip}%
}%
\item[]}{\end{list}}

\begin{document}
\begin{titlepage}

\begin{changemargin}{-2cm}{-2cm}
\vspace{.3in}
\hfill{ \begin{flushright}UMD-PP-013-013\\ CERN-PH-TH-2014-220 \end{flushright}}
\vspace{2.0cm}
\openup 0.8em
\begin{center}
\noindent \textbf{\textsc{\Large\textcolor{RoyalBlue}
{Using Energy Peaks to Measure New Particle Masses}}
}
\end{center}
\openup -0.8em

\begin{center}
\vspace{.75cm}
{\bf Kaustubh Agashe\symbolfootnote[1]{\href{mailto:kagashe@umd.edu}
{kagashe@umd.edu} }, 
Roberto Franceschini\symbolfootnote[2]{\href{mailto:rfrances@umd.edu}
{rfrances@umd.edu} }, 
%
%
Doojin Kim\symbolfootnote[3]{\href{mailto:immworry@umd.edu}
{immworry@umd.edu} }\symbolfootnote[4]{
Address after September 1, 2013: Institute for Fundamental Theory, Physics Department,
University of Florida, Gainesville, FL 32611, U.~S.~A.} }\\
\setstretch{1.5}
Maryland Center for Fundamental Physics, Department of Physics,\\
University of Maryland, College Park, MD 20742, U.~S.~A.
\setstretch{1.}
\end{center}

\end{changemargin}


\begin{abstract}
\medskip
\noindent
We discussed in arXiv:1209.0772 that the laboratory frame distribution of the energy of a massless particle from a two-body decay at a hadron collider
has a {\em peak} whose location is identical to the value of this daughter's (fixed) 
energy in the {\em rest} frame of the
corresponding mother particle. For that result to hold we assumed that the mother is {\it un}polarized and has a {\it generic} boost distribution in the laboratory frame.
In this work we discuss how this observation can be applied for determination of masses of new particles, with{\em out} requiring a
full reconstruction of their decay chains or information about the rest of the event. 
We focus 
%
%
on a two-step cascade decay of a massive particle that has one
%
%
invisible particle in the final state:
$C \rightarrow Bb \rightarrow Aab$, where $C$, $B$ and $A$ are new particles of which $A$ is invisible and $a$, $b$ are visible particles. Combining the measurements of the peaks
of energy distributions
of $a$ and $b$ with that of the edge in their invariant mass distribution, we demonstrate that it is in principle possible to determine separately {\it all}
three masses of the new particles, in particular, without using any measurement of missing transverse momentum. 
Furthermore, we show how the use of the peaks in an inclusive energy distribution is generically less affected (as compared to other mass measurement strategies) by combinatorial issues. 
For some simplified, yet interesting, scenarios we find that these combinatorial issues are absent altogether.
As an example of this general strategy, we study SUSY models where gluino decays to an invisible lightest neutralino via an on-shell bottom squark. Taking into account the dominant backgrounds, we
show how the mass of the bottom squark, the gluino  {\em and} (for some class of spectra) that of the neutralino can be determined using this technique.
\end{abstract}

\end{titlepage}


\setcounter{tocdepth}{2}
\tableofcontents

\section{Introduction}

The Large Hadron Collider (LHC) has a great potential 
to discover an extension of the Standard Model (SM) at the TeV scale.
Such new physics is especially motivated by solving the Planck-weak hierarchy problem of the SM.
Furthermore, a (stable) weakly interacting massive particle (WIMP), with a weak-scale mass, is an attractive
candidate for the Dark Matter (DM) of the universe since its thermal freeze-out can give the correct relic density.

Once produced at the LHC, such new particles  are likely to decay into SM particles (since they arise in an extension thereof).
In some cases, some of the new particles are charged under a new symmetry, while SM particles are not. Thus, the lightest new particle (LNP) is stable and can be DM if it is colorless and electrically neutral. In such a scenario the heavier new particles decay 
into SM particles and an invisible LNP.
As a corollary, such new particles cannot be singly produced and are instead usually produced in pairs.

All in all, a signal for the discovery of such new particles would come from an observation of excess with respect to the SM prediction of final
states with SM states and  missing momentum.
Several techniques, based on the underlying kinematics of such processes, have been suggested and used {\it recently} for effectively carrying out such searches: for example, the variables $M_{T2}$ and its variations~\cite{Lester:1999et, Barr:2003fj, Cho:2007qv, Barr:2011ao},
$\alpha_T$~\cite{Randall:2008rw} or razor ~\cite{Rogan:2010kb} (for a review, 
see, for example, Ref.~\cite{Gripaios:2011kk}).
Once discovered, the next stage of investigation would obviously be the determination of what is the extension of the SM at play 
in this signal.
%
%
Eventually, we would like to pin down the specific model that is realized in Nature within a more general framework, which could be %
%
for example
a particular model of SUSY or composite Higgs. 
%
%
%
%
It is clear that to achieve the above goal we need to probe the properties of the new particle such as spin, mass, couplings, electroweak charge, color etc.
In turn, for this purpose, we may need to reconstruct the decay chain of the new particle.

In this work we focus on  the measurement of the {\em mass}  of the new particles.
If the new particle decays only to SM visible particles, then this task might seem ``easy'' since
constructing the invariant mass of the decay products provides, in principle, the full information on the mass of the new particle~\footnote{
Here, we do not consider the possibility that the new particle decays into a final state that has some neutrinos, which would make the use of invariant mass not possible.}.
However, often this method is plagued by combinatorial issues, since it is possible that there is 
more than one new particle in  each event or the new particle is produced in association with some other SM visible particles.
In this case, a priori we 
do not know which visible particles came from a given new particle.
On the other hand, even assuming that the correct grouping of the visible particles into candidate resonances can be achieved, there are some models where this is still not enough because the final state of the decay of the new particle just does not contain enough information to fully reconstruct the invariant mass of each of the mother particles. For example, in some models the new particle decays to SM and invisible LNP as discussed above.
It is apparent that in this case the invariant mass of the visible particles from the decay of the new particle gives some 
{\em combination} of  the masses of the new particles, but typically can{\em not}
provide the information about each mass separately. 

In order to get the new particle masses separately, we might then have to use the missing momentum carried away by LNP, as it is the case for most of the existing techniques, especially for short decay chains. 
Since the new particles are pair-produced in this case with each new particle decaying into LNP, the missing momentum
is shared
between the two new particle decays. Thus, it seems that if we use the missing momentum to infer the properties of each single new particle, we are bound  to make use of the full information about the event. In other words, the fact that the observed missing momentum is the result of two momenta belonging to particles from different decay chains inevitably entangles the study of one decay chain to the other.
For such cases,
the $M_{ T2}$ variable (and related 
%
%
ones) \cite{Lester:1999et, Barr:2003fj, Cho:2007qv, Barr:2011ao}
have been designed to determine the individual new particle masses along these lines~\footnote{See also, for example,
Refs.~\cite{Cheng:2007xv,Han:2009ss,Cho:2012gd} for other recent methods of mass measurement that do not use missing transverse energy and Refs.~\cite{Barr:2010hs,Gripaios:2011kk} for a general review of mass measurement methods. }.
In order to give 
%
%
precise measurements of the new particle masses, these analyses typically require 
(in turn)
accurate measurements of 
the missing momentum, which can be
%
%
plagued by large uncertainties.
Furthermore, they can usually be applied only to the cases where the two decay chains involve the same new particles.
Thus, we are motivated to develop new strategies that can deal with the decays of new particles involving invisible particles in
order to possibly get around the above mentioned limitations.
In particular, our goals are {\it a}) to avoid the use of the missing momentum, which is poorly measured; {\it b}) to be able to deal with generic production mechanism of the new particles, not just with the case of productions in pairs; {\it c}) and to be as safe as possible from the issues that arise from combinatorics. 
Such new techniques can certainly be complementary to (if not better than) known ones.

With the above motivations in mind, in an earlier paper of ours~\cite{Agashe:2012mc},
the energy spectrum of a massless product from a two-body decay of a massive particle at hadron colliders was studied.
In particular, we have been able to show that the typical conditions for the production of massive particles at  hadron
colliders are such that, in the two-body decay of a massive particle, the energy distribution of a massless daughter particle in the laboratory frame has a {\em peak} located at the same value as the corresponding energy which would have been seen in the rest frame of such a massive decaying particle (in this sense, we henceforth call it ``rest-frame energy'' of the 
daughter particle)~\footnote{After our work~\cite{Agashe:2012mc} was submitted, we found that this basic result on which our techniques for mass measurement rely on had appeared in previous work~\cite{1971NASSP.249.....S} about cosmic ray physics. We remark that our results are more general than those of Ref.~\cite{1971NASSP.249.....S}, which dealt only with the case of scalar decaying particles (i.e., $\pi^0$). In fact, our results, which are recalled later in Section~\ref{theory}, also cover the case of particles with spin. We stress that, to the best of our knowledge, the observation made in Ref.~\cite{1971NASSP.249.....S} (and~\cite{Agashe:2012mc}) has not been applied previously in high-energy particle physics.}.
This means that this peak retains the information about the masses involved in the decay in a way that is as simple and precise as the information inferred from decay kinematics in the rest frame of the decaying particle. In order for this result to hold we required  that 
%
%
the mother particles in the event sample under study are produced unpolarized along with a distribution of boosts which includes
small values (see the next section for the precise version of this condition).
We stress that these assumptions are rather generically satisfied at high energy hadron colliders.

It is clear that the above observation can be useful to determine the mass of the mother particle undergoing a two-body decay, provided we can extract this energy peak accurately from data.
In fact, we showed earlier~\cite{Agashe:2012mc} that the mass of the top quark can be rather accurately extracted using the energy peak, i.e., from the $b$-jet energy spectrum arising from the  decay $t\to bW$. 
As part of this study, we also proposed a fitting function for the energy spectrum that has been proven to work extremely well to extract the peak position from the data. Armed with this fitting function, we are then ready to discuss applications of the above observation to spectroscopy
of {\em new} particles which we expect to be discovered at the LHC. 
Another part of the goal in the present paper is the following.
Although our measurement of the top quark mass in~\cite{Agashe:2012mc} was performed including detector effects, soft QCD radiation, and the necessary event selection to isolate top quark events, 
%
%
we concentrated there on the features of the signal process, and hence backgrounds were not considered.  
Thus, in the present study of mass measurement of new particles, we aim 
to amend for this earlier simplified treatment of the backgrounds by showing that energy peaks can be useful even in the presence of backgrounds.

It is 
remarkable
 that with the  
 %
 %
 technique first presented in~\cite{Agashe:2012mc} we can 
 %
%
measure
masses without any recourse either to the full reconstruction of the decay chains or to the global information, i.e., knowledge of 
the rest of the event. In this sense, the use of energy peaks relies only on the information from the subset of the event that arises from the decay of interest, and thus it exploits only ``local'' information in the event. The ``locality'' of our technique brings significant advantages in the situations where the global topology of the event is not known, either because of our ignorance on the source of the events, or because of an inclusive treatment of different sources for the decay under consideration. Furthermore, we stress that our technique relies on the measurement of peaks, which are intrinsically easy features to spot in the relevant distributions, rather than the often more difficult to measure endpoints, which are usually the subject of other techniques (for instance, those based on $M_{T2}$ or invariant mass variables).
For completeness of the discussion, we remark that our basic mass measurement technique (outlined above) is somewhat related to that of Refs.~\cite{Kawabata:2011gz,Kawabata:2013fta}. 
In fact, the starting point for both our technique and that of Refs.~\cite{Kawabata:2011gz,Kawabata:2013fta} is to make use of 
the {\em energy} spectrum (i.e., a quantity which is manifestly not Lorentz-invariant), that too of only {\em one} decay product, yet come up with an observable which is not dependent on the boost of the mother particle. 
However, the two techniques differ in several ways. The authors of Ref.~\cite{Kawabata:2011gz,Kawabata:2013fta} achieve this goal of insensitivity to the mother particle boost by constructing an observable that is a kind of a weighted average of the energy of the daughter particle. Interestingly, their results hold for an {\em arbitrary} distribution of the boost of the mother particles. However, 
achieving such robustness necessitates utilizing the {\em entire} energy distribution, i.e., including the tails. 
Our technique instead relies on a local feature of the energy distribution, i.e. the peak, which usually can be efficiently searched without knowing the rest of the distribution and might be more readily identified in the data even when it appears over backgrounds. We recall that our result does not hold for a {\em completely} arbitrary distribution of the boost of the mother particle, however, as we have mentioned above, the assumptions about boost distribution needed for our result to hold are quite generically realized at hadron colliders.

In this paper we carry out the study of a  semi-invisible two-step cascade decay of a 
new particle: 
\beq
C \rightarrow B\,b \rightarrow A\,a\,b,\label{genericdecay}
\eeq
where $C$, $B$ and $A$ are 
(on-shell) new particles of which $A$ is unobserved and $a$, $b$ are visible (SM) particles. 
According to our above observation, the peak in the observed energy distribution of the $b$ gives a relation between $C$ and $B$ masses. Similarly, the peak of the observed energy distribution of $a$ relates masses of $B$ and $A$. Finally, 
it is well-known \cite{Hinchliffe:1999ve,Allanach:2000gf} that the invariant mass of $a$ and $b$ has an 
edge that gives a relation between all three masses, which we show later is independent of the first two relations.
The existence of these three relations
implies that we can determine all the three  masses  of the new particles  
$C$, $B$ and $A$.
We stress that the mass measurement strategy proposed in this paper 
%
%
involves 
neither 
the measurement of the missing transverse momentum 
nor 
%
%
the information from the rest of the event, in particular, 
%
%
about any new particles produced ``on the other side''.
Once again we want to contrast these features with those of the methods based on $M_{ T2 }$-type variables which
require missing momentum measurement and usually can be applied only when the initiator of the decay chain is the same for both sides.

Another potentially attractive feature of our method (compared to the existing ones)  is as follows. 
As already mentioned, mother particles in the scenario that we are considering are typically produced in pairs.
Suppose that each mother particle decays into more than one visible particle (as in \eq{genericdecay} above) and furthermore undergoes the same decay on both ``sides'' of the event.
Most of the existing observables such as the ones based on $M_{T2}$ or invariant mass of visible particles
%
%
require ``combining" momenta of the particles which originate from the same decay chain.
Clearly, for such observables, it is necessary to identify correctly which side of the event the relevant decay products
came from.
Several strategies for such correct partitioning of the final state particles have been suggested~\cite{Albrow:1976jm,Hinchliffe:1997fk,Dutta:2011uq,Rajaraman:2011fj,Choi:2011lr,Curtin:2011ng}.  
We demonstrate in the following that for such types of production mechanisms and decay patterns of the new particles
our method is largely  unaffected by these combinatorial issues. 
This implies that no special strategies for resolving such issues need to be developed to apply our technique. This is rather relieving because the strategies that can be applied to alleviate the above mentioned combinatorial issues often rely on educated guesses on the preferred  kinematics of the underlying signal, and therefore, they usually need to be validated with care for each case.

In this work, we apply our above general strategy to a {\em specific} example of Eq.~(\ref{genericdecay}) from SUSY, namely,  gluino pair-production with 
each gluino decaying via on-shell bottom squark to lightest neutralino (which is invisible) and two $b$-jets:
\beq
pp\to \tilde{g}\tilde{g} \to bb\tilde{b}\tilde{b}\to 4b\met\,.
\label{gluino}
\eeq
As per the observation above, we will get {\em two} peaks in the observed $b$-jet energy
distribution, corresponding to the two two-body decays, i.e. the one from a gluino decay and the one from a sbottom decay. 
This type of event is completely symmetric, i.e., it has two identical decay chains.
%
%
In spite of this feature, it is clear that 
in this case our strategy does not suffer any problem from combinatorial issues in the peak analysis. 
In fact, in general, for identical decay chains of the type in \eq{genericdecay}, there are two $a$ particles in  each event, but both are from the two $C$ particles obeying the decay described in Eq.~(\ref{genericdecay}). In a similar manner, we have two $b$ particles per event. However, the presence of several identical particles in the final state does not affect our study of the energy distributions because the result of Ref.~\cite{Agashe:2012mc} can be applied independently for each single two-body decay in the process. This must be contrasted with the invariant mass of any combination of $a$ and $b$ particles where we might (wrongly) pair $a$ from one $C$ with $b$ from the other $C$. The benefit against combinatorics that arises from the ``locality'' of our observable is at full display in the specific example that we have chosen to study. In fact, in our example not only there are two $a$ and two $b$ particles per event, but we even take $a$ and $b$ to be identical, thus giving {\it four} identical particles,  a situation which would of course maximally complicate the combinatorial issues that afflict {\em other} methods.

In this study we analyze two types of  spectra. The first one has gluino and sbottom close in mass, but both being much heavier than neutralino. 
In this case  the two peaks in the $b$-jet energy distribution are
well-separated such that they can be seen in the (simulated) {\em signal}
data even ``by eye''. 
In spite of the lower peak being below $\sim100$ GeV, where QCD backgrounds
might be large, it can still be extracted from the {\em combined} (i.e., signal+background) distribution {\em above} $\sim100$ GeV
by the fitting procedure mentioned above for the case of top quark mass
and described in detail later. 
In spite of measuring three
combinations of the three a priori unknown masses, we find that there is not much sensitivity to neutralino mass for this kind of spectrum. The lack of sensitivity to the mass of the neutralino is somewhat expected since this mass is negligible with respect to the gluino and sbottom masses, thus having in general a very modest impact on the kinematics of the events. However, we stress that with our technique, enough information can be extracted from the events so that we can determine at least the gluino and sbottom masses separately.

The second type of spectrum that we study has a gluino significantly heavier than {\em both} sbottom and neutralino. In this case we find that the two-peak structure cannot be seen in the data by eye. However, once again the fitting procedure introduced in~\cite{Agashe:2012mc} is powerful enough to find {\em both} peaks. In this case the mass of the neutralino is not negligible compared to the other masses and we can 
have  sensitivity to the neutralino mass as well as the sbottom and gluino masses.

We emphasize that for other (than \eq{gluino}) types of decay chains, 
the technique can be even more successful than in the above example. For instance, if $a$ and $b$ in Eq.~(\ref{genericdecay}) are not identical, the fitting procedure for extracting the peaks from the data is much more straightforward and robust, and thus typically gives significantly smaller errors on the masses since the relevant energy-peaks appear in {\em different} distributions (unlike the case for the process in \eq{gluino}).
Furthermore, we expect that when the decay chain involves fewer jets and more leptons the new particles masses can be ever more accurately extracted. 
The reason is that lepton energies can be measured more precisely than jets
and 
the contamination of signal due to initial state radiation 
%
%
is also reduced.
Additionally, when either $a$ or $b$ in the decay Eq.~(\ref{genericdecay}) are leptons the threshold on their $p_T$ can be lower than that used for jets so that the extractions of peaks at low energy can be more easily carried out.
%
%
Finally,
we believe that it is certainly possible to combine our observation about the peak in the observed energy distribution(s) with other techniques (like we already do in this work by combining it with the invariant mass edge)
in order to develop even better techniques for mass measurement. At the very least, we believe that 
our observation can be used as a ``cross-check'' for other techniques.

\bigskip

The rest of this paper is organized as follows. We begin in Section~\ref{theory} by 
reviewing the central observation about the peak in the observed energy spectrum of a massless
daughter from two-body decay. 
We then move onto our main focus in Section~\ref{mainsection}, namely, the application to a semi-invisible, two-step cascade decay of new particles. In Section~\ref{susygluino}, 
we carry out explicitly the study of the gluino/sbottom  example from SUSY. Finally, in Section~\ref{conclusion} we give our conclusions  and and outlook on future work.

\section{Laboratory frame energy distribution from a two-body decay}
\label{theory}

In this section we review results from our earlier work~\cite{Agashe:2012mc} which will be used in the following sections.
In Ref.~\cite{Agashe:2012mc} we investigate the connection between the energy of the visible daughter in the {\it rest} frame of the associated mother particle and the position of the {\it peak} in its energy distribution seen in the laboratory frame for two-body decays. We consider the decay of a heavy mother particle $B$ into a {\em massless} visible particle $a$ along with a massive invisible particle $A$:
\bea
B \rightarrow A\,a.
\eea
In what follows  all of the properties of the particle $A$ will be irrelevant, but for its mass that we denote it as $m_{A}$. We stress that the argument that we are going to recall from Ref.~\cite{Agashe:2012mc} is valid for any value of $m_{A}$ allowed by the decay of the particle $B$, including the case $m_{A}=0$, where our results would get even simpler.

It is well-known that the energy of the visible particle in the rest frame of the mother particle can be expressed in terms of two mass parameters $m_B$ and $m_A$:
\bea
E^*=\frac{m_B^2-m_A^2}{2m_B}.
\eea
Here and henceforth the superscript `$\ast$' implies that the associated quantity is measured in the rest frame of the corresponding mother, i.e., here particle $B$. 
$E^*$ is trivially single-valued so that the shape of the energy distribution in the rest frame appears as a $\delta$-function. However, once the mother particle, or, equivalently, the overall system,  is boosted by a (fixed) Lorentz factor $\gamma \equiv1/\sqrt{1-\beta^2}$ from the rest frame to the laboratory frame the ``spiky distribution'' of the rest frame gets flattened out and in fact is given by:
\bea
E=E^*\gamma\left(1+\beta\cos\theta^*\right)\,, \label{eq:Elab}
\eea  
where $\theta^*$ defines the intersecting angle between the direction of emission of the particle $a$ and the boost direction $\vec{\beta}$. 
If the mother particle has  spin-$0$, by definition there are no preferred directions in its decay and the $\cos\theta^{*}$ distribution is flat. On the contrary, if a particle has spin, the distribution of $\cos \theta^*$ in general is {\em not} flat, because its spin defines a preferred direction in space. However, it is possible that the mother particle is produced by an interaction, for instance strong or electromagnetic interactions, which do not distinguish the different states of polarization. 
In this case, the symmetries of the interactions responsible for the production mechanism guarantee that a given set of particles is {\em effectively} produced {\em unpolarized}, i.e., 
%
%
the spin direction of the mother particle with respect to the boost direction in the rest frame takes all possible values with equal probability, hence the $\cos \theta^*$ distribution is flat.
The flatness of the $\cos\theta^{*}$ distribution implies that the  $E$ distribution should be a flat  as well. More precisely, since $\cos\theta^*$ runs from $-1$ to $+1$, for any given $\gamma$ the shape of the distribution in $E$ is simply given by a ``rectangle'' covering the range 
\bea
E\in \left[E^*\left(\gamma-\sqrt{\gamma^2-1}\right),E^*\left(\gamma+\sqrt{\gamma^2-1}\right)\right]\,. \label{eq:rangeofE} 
\eea

In order to obtain the energy distribution in the laboratory frame, for any given $E$ the contributions from all relevant $\gamma$ factors must be superimposed. Considering the fact that the shape of $E$ for every single $\gamma$ is a simple rectangle covering the range of Eq.~(\ref{eq:rangeofE}), the superposition mentioned before can be understood as ``stacking up'' many such rectangles~\footnote{As it will be explained in greater details later, the relative heights  of the rectangles is determined by the actual distribution of the boost of the mother particle. However, this detail has no impact at all on our statement about the peak position in the laboratory frame energy distribution.}. 
One crucial observation is that 
each contribution
coming from a fixed, but arbitrary $\gamma$ has common support on the value $E=E^*$. This is clear from the fact that the lower (upper) bound in Eq.~(\ref{eq:rangeofE}) is less (larger) than $E^*$ and that the distribution is flat in-between. Remarkably, there is no other value of $E$ which attains this feature as far as the $\gamma$ distribution is non-vanishing in a small region around $\gamma=1$. Moreover, for fixed $\gamma$, due to the rectangular shape of the distribution, it is clear that no other value of $E$ gets a larger contribution than  $E^*$ does. Therefore, the distribution of $E$ has a peak  manifestly located at $E=E^*$. Denoting the distribution of the laboratory frame energy $E$ as $f(E)$ we state our finding by the simple equation:
\bea
f_{\max}=f(E^*).
\eea
We emphasize that it is {\it not} obvious at all that the peak in the energy distribution of the visible daughter is identical to its rest-frame energy, especially for the decay of particles with spin. 

Another noteworthy feature of the laboratory energy distribution is that it is {\em not} symmetric with respect to the peak position at $E=E^*$. This can be seen from the fact that  the distance of the upper bound in Eq.~(\ref{eq:rangeofE}) from the peak is farther than that of the lower bound. Thus, the energy distribution of particle $a$ develops a longer tail toward high energy with respect to the peak. 

We can understand the existence of the peak with a more formal derivation. We recall the fact that $\cos\theta^*$ is a flatly distributed variable, which implies that the differential decay width in $\cos\theta^{*}$ is constant:
\bea
\frac{1}{\Gamma}\frac{d\Gamma}{d\cos\theta^*}=\frac{1}{2}\,.
\eea
From this simple relationship along with Eq.~(\ref{eq:Elab}) one can easily derive the differential decay width in $E$ for any fixed boost factor $\gamma$:\footnotesize 
\bea
\tiny
\left.\frac{1}{\Gamma}\frac{d\Gamma}{dE}\right|_{\mathrm{fixed}\;\gamma}&=&\left.\frac{1}{\Gamma}\frac{d\Gamma}{d\cos\theta^*}\frac{d\cos\theta^*}{dE}\right|_{\mathrm{fixed}\;\gamma} \nonumber \\
&=&\frac{1}{2E^*\sqrt{\gamma^2-1}}\Theta\left[E-E^*\left(\gamma-\sqrt{\gamma^2-1}\right) \right]\Theta\left[-E+E^*\left(\gamma+\sqrt{\gamma^2-1}\right) \right] \normalsize
\eea \normalsize
where the two $\Theta(E)$ are the usual Heaviside step functions which merely constrain the allowed region of $E$. In order to have the complete expression for any given $E$, we have to integrate over all $\gamma$ factors which affect the value of $E$ that we are interested in. Denoting the probability distribution of $\gamma$ by $g(\gamma)$, the normalized energy distribution $f(E)$ can be cast into the integral form:
\bea
f(E)=\int_{\frac{1}{2}\left(\frac{E}{E^*}+\frac{E^*}{E} \right)}^{\infty}d\gamma \frac{g(\gamma)}{2E^*\sqrt{\gamma^2-1}}\label{eq:f}\,.
\eea
The lower end in the integral comes from solving the equation for  the minimal $\gamma$ factor that can affect a given energy:
\bea
E=E^*\left(\gamma\pm\sqrt{\gamma^2-1}\right)
\eea
with the positive (negative) sign being relevant for $E\geq E^*$ $(E<E^*)$. We can  also compute the first derivative of Eq.~(\ref{eq:f}) with respect to $E$, which is:
\bea
f'(E)=-\frac{1}{2E^*E}\mathrm{sgn}\left(\frac{E}{E^*}-\frac{E^*}{E} \right)g\left(\frac{1}{2}\left(\frac{E}{E^*}+\frac{E^*}{E} \right) \right).\label{eq:fprime}
\eea

The solution of $f'(E)=0$ are the extremal points of $f(E)$, which, as shown by Eq.~(\ref{eq:fprime}), are coming from the zeroes of the mother particle boost distribution $g(\gamma)$. Typically, for particles produced at colliders the boost probability distribution $g(\gamma)$ is not vanishing in a range of $\gamma$ from 1 to some upper limit set by the energy available at each collider. Therefore, as far as zeros are concerned two possible cases can arise here depending on whether or not $g(1)$ (corresponding to $E=E^*$) vanishes. If it vanishes, then $f'(E=E^*) \propto g(1)=0$, which implies that the distribution has its unique maximum point at $E=E^*$~\footnote{This particular result was also found in Ref.~\cite{1971NASSP.249.....S}, as mentioned above.}. If $g(1)\neq 0$, then $f'(E)$ flips its  sign at the point $E=E^*$ due to the presence of the sign function in Eq.~(\ref{eq:fprime}). As a result, the distribution shows a cusp concave structure at $E^{*}$, which is still giving a peak in the energy distribution at $E^*$.

As explicitly discussed in the Introduction, in order to apply this observation to mass measurement  it is essential to accurately extract the location of the peak  from data. However, having the analytic expression for the shape of the energy distribution $f(E)$ only relying on first principles seems very difficult. The reason is that the details of the boost distribution $g(\gamma)$ are sensitive to the internal structure of the protons or any other initial state of the collider,  the mother particle mass, and the actual decay vertex of the mother particle. Nevertheless, there are some functional properties which the energy distribution $f(E)$ should obey. We list some of them below:
\begin{enumerate}
\item $f$ is a function over an argument of $\frac{1}{2}\left(\frac{E}{E^*}+\frac{E^*}{E} \right)$, i.e., it is even under  $\frac{E}{E^*}\leftrightarrow\frac{E^*}{E}$,
\item $f$ has a global maximum at $E=E^*$,
\item $f$ vanishes when $E \rightarrow 0$ or $E\to\infty$,
\item $f$ becomes a $\delta$-function for some limiting parameter choice.
\end{enumerate}
Based on these properties, in Ref.~\cite{Agashe:2012mc} we proposed a well-motivated {\it ansatz} for the functional form of the energy distribution in the laboratory frame:
\bea
f(E)=K_1(w)^{-1}\exp \left[-\frac{w}{2} \left (\frac{E}{E^*}+\frac{E^*}{E} \right) \right] \label{eq:fitter}\,,
\eea
where $w$ is a fitting parameter affecting the width  of the peak, and the normalization factor is given by a modified Bessel function of the second kind of order 1, $K_{1}(w)^{-1}$. Clearly, our ansatz for $f(E)$ fulfills all the properties listed above. In Ref.~\cite{Agashe:2012mc} it was  shown that $E^*$ can be extracted by fitting the data of interest with Eq.~(\ref{eq:fitter}). In particular, the determination of $E^{*}$ by mean of our ansatz was tested on the decay $t\to bW$. In that study a very good agreement between the function and the data was observed: a very good agreement in the peak region and a slightly less good agreement in the tails.
From these results it is clear that the ansatz Eq.~(\ref{eq:fitter}) has very good chances to give good results also for the energy distribution that we will study in the following sections.

\section{Application to two-steps decay chains: General Strategy \label{mainsection}}
In this section we demonstrate how the observation described in the preceding section can be used for mass measurement in a realistic setting, which includes contaminations from backgrounds as well as a treatment of the mis-modeling of the signals away from the peak region. For this purpose we employ a two-step cascade decay as a concrete example~\footnote{Even if we take the two-step cascade decay, we stress that the idea can be easily extended to the generic multi-step cascade decays.}:
\bea
C\rightarrow B\,b \rightarrow A\,a\,b \label{genericdecayofC}\,,
\eea
where particle $A$ is assumed to be invisible, particle $B$ is assumed {\it on}-shell, and particles $a$ and $b$ are assumed visible and massless.

For the process of interest there are three unknown mass parameters: $m_A$, $m_B$, and $m_C$. If we find three independent relations among the three masses and some observables, then we can in principle determine all these masses. Our general strategy is to take two relations from the peaks in the energy distributions of the two visible particles, which is the main novel ingredient in the mass measurement strategy that we discuss in our work. According to the discussion in Section~\ref{theory}, they  are given by:
\bea
\Eapeak=\frac{m_B^2-m_A^2}{2m_B}\label{eq:peaka}\,, \\
\Ebpeak=\frac{m_C^2-m_B^2}{2m_C}\,.\label{eq:peakb}
\eea 
Another observable that we consider in order to get a third independent relation is the well-known kinematic endpoint in the invariant mass distribution formed by the two visible particles \cite{Hinchliffe:1999ve,Allanach:2000gf}~\footnote{Of course, the use of this relation is simply incidental as we need a third relation to close the system of equations. Using this particular observable is merely an option, and not at all necessary to illustrate our point about the usefulness of using energy peaks. In fact, we could use any other equation that relates  some observable to the masses,  as long as it provides an independent equation.}:
\bea
m_{ab,\max}=\sqrt{\frac{m_C^2-m_B^2}{m_B}\,\,\frac{m_B^2-m_A^2}{m_B}}=2\sqrt{\frac{m_C}{m_B}\Ebpeak \Eapeak }\,.\label{eq:invedge}
\eea
Inverting  Eqs.~(\ref{eq:peaka})-(\ref{eq:invedge}) we obtain expressions for the three mass parameters in terms of the three observables  $\Eapeak$, $\Ebpeak$, and $m_{ab,\max}$:
\bea
m_C&=&\frac{2m^{4}_{ab,\max} \Ebpeak}{ m^{4}_{ab,\max}-16 \Ebpeak^{2} \Eapeak^2 }\,, \nonumber \\
m_B&=&\frac{8 m^{2}_{ab,\max} \Ebpeak^{2} \Eapeak }{m^{4}_{ab,\max}-16 \Ebpeak^{2} \Eapeak^2 }\,, \label{massesfromobservables}
 \\
m_A&=&\frac{m_{ab,\max} \Ebpeak \Eapeak} { \frac{m^{4}_{ab,\max}}{16}-\Ebpeak^2 \Eapeak^2 }\sqrt{\Ebpeak^2\left(\frac{m^{2}_{ab,\max}}{4}+\Eapeak^2\right) -\frac{m^{4}_{ab,\max}}{16}}\,. \nonumber
\eea

These equations fully display the advantages of using energy peaks to measure particle masses.
In fact, the masses (including that of the {\em invisible} particle!) can be obtained from observables that do not depend at all on missing transverse momentum, which is rather difficult to measure accurately. 
Furthermore, most of the quantities used to compute the masses are extracted from single-particle observables, which means that there we greatly reduce the combinatorial issues that arise from the formation of  multi-particle systems.

It is useful to remark some inequalities that must hold for these inverse formulae to be applied. Since the numerators for $m_B$ and $m_C$ are already positive, the denominators must be positive as well for getting them physical, and thus we have:
\bea
m^{2}_{ab,\max}> 4 \Eapeak \Ebpeak\,.
\eea
Furthermore, the expression inside the square root in the expression for $m_A$ should be positive. Therefore, $m^{2}_{ab,\max}$ acquires the upper bound:
\bea
m^{2}_{ab,\max} < 2\Ebpeak \left(\Ebpeak+\sqrt{\Ebpeak^{2}+4 \Eapeak^{2}} \right).
\eea
All in all we see that the three observables must satisfy the following hierarchy:
\bea
4 \Ebpeak \Eapeak <m^{2}_{ab,\max} <  2 \Ebpeak \left(\Ebpeak +\sqrt{\Ebpeak^{2}+4 \Eapeak^{2}} \right). \label{inequalities}
\eea

{In the above discussion we have assumed that one is able to say what two-body decay originate each peak in the energy distribution. Without any a priori knowledge of the mass spectrum we may not be able to assign correctly the energy peak to a decay step in the chain and in general the assignment of each peak to a decay step should be questioned. When one observes the two energy peaks of a two-step decay chain as that in \eq{genericdecayofC} two possible assignments for the energies are available. These two assignments are equivalent to swap $\Eapeak$ and $\Ebpeak$ in \eqs{eq:peaka}-(\ref{eq:invedge}) and in general give different results for the reconstructed masses. We note in \eqs{massesfromobservables} that the masses in terms of the observables  are such that they can be schematically written as
\beq m_{A}=\mathcal{S}_{A}^{(ab)}\cdot \Ebpeak\,\quad m_{B}=\mathcal{S}_{B}^{(ab)}\cdot \Ebpeak\,\quad m_{C}=\mathcal{S}_{C}^{(ab)}\cdot\sqrt{\tilde{\mathcal{S}}_{C}^{(ab)}+\Ebpeak}
\eeq
where the various $\mathcal{S}^{(ab)}$ are each a symmetrical expression under the exchange of $\Eapeak$ and $\Ebpeak$. From this schematic rewriting of \eqs{massesfromobservables} we can conclude that  assigning $\Ebpeak$ to the largest or to the smallest of the measured energy peaks induces an overall change in the spectrum by a factor of order $\Eapeak/\Ebpeak$. For many cases, as the ones that we discuss in the following, $\Eapeak/\Ebpeak\simeq few$, therefore the two possible interpretations of the energy peaks return spectra of significantly different overall mass scale. In fact we estimate that for more than 20\% difference in the measured energy peaks, i.e.  $\left| \Eapeak/\Ebpeak - 1 \right|>0.2$, the two possible spectra can be distinguished just looking at the cross-section of the signal, that for masses differing more than 20\% should be one order of magnitude or more different. When the energy peaks become closer  cross-section considerations are no longer so helpful and one should attempt both assignments of the energy peaks. However we remark that as one get closer to $\Eapeak=\Ebpeak$ the two spectra will coincide, and the whole discussion of the assignments of the energy peaks becomes empty of meaning. Furthermore, before hitting the case of equal energy peaks, one should start to question about a number of experimental issues, such as the energy resolution that, for instance for jets, could just not be enough to distinguish energy peaks that differ by order 10\%. From this brief discussion we conclude that in most cases it is quite easy to build up confidence about what is the correct assignment of the energy peaks and that an ambiguity remains only when the two option for the spectrum between which we are called to choose are essentially the same. 
On top of the cross-analysis of rates and energy and $m_{bb}$ spectrum one can add further elements to decide how to assign the energy peaks to the two-body decays in the chain. We have seen that the internal consistency of the system of equations that we need to invert to obtain the masses from the observables implies the inequalities \eq{inequalities}. In some cases, such as the case that we dicuss in Section~\ref{spectrumone},  these inequalities just are not satisfied if one swaps $\Eapeak$ and $\Ebpeak$, thus leaving only one possible interpretation of the energy peaks.\footnote{An alternative way to try to determine the part of the cascade decay chain to which each of the two peaks correspond is to study the fitted width parameter ($w$) since this value depends on the boost of the associated decaying particle, which (in general) is different for the primary and secondary mothers. However, we find that the extracted width also depends significantly on details such as fitting procedure and event selection. Therefore, we think that in the end the use of the fitted width parameter would be limited to corroborating evidence that already arose from the cross-section considerations mentioned above. }
Given the several checks described above that one can use to assign the energy peaks to the two-body decays, in what follows we do not consider the issue of the assignment of the energy peak any more and we simply assume that the correct assignment has been understood before trying to reconstruct the masses.
}

\section{Application to gluino decay in SUSY\label{susygluino}}
\begin{figure}[t!]
\begin{center}
\includegraphics[width=1.0\linewidth]{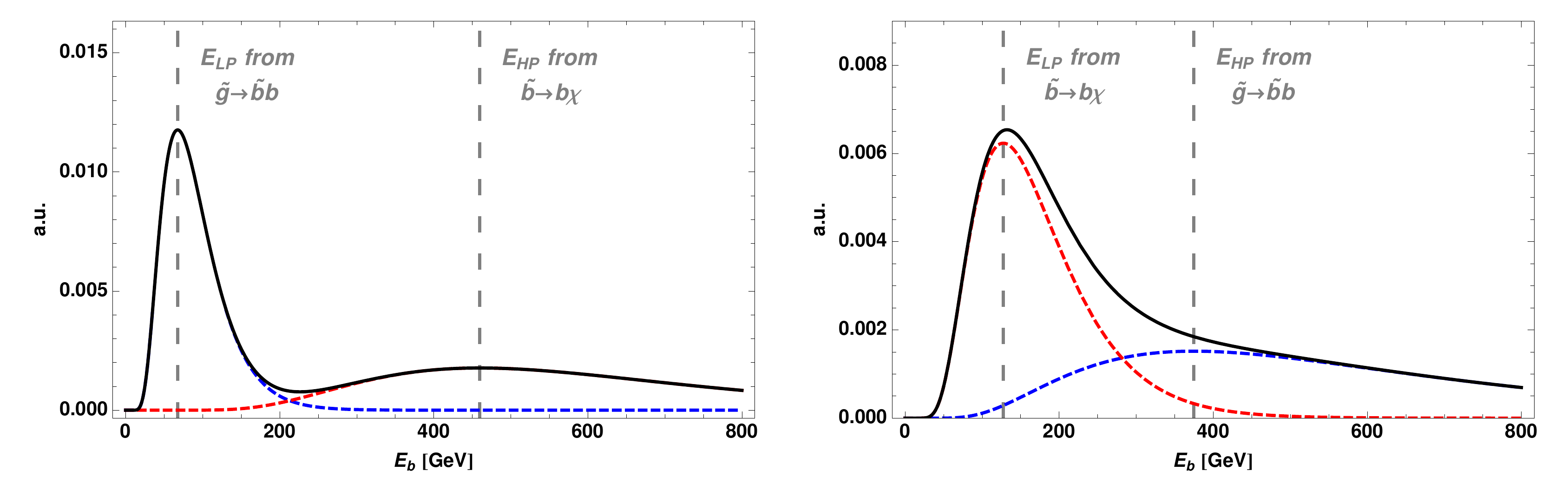}
\caption{A schematic decomposition of the $b$-jet energy spectrum that we expect for mass spectra with degenerate gluino and sbottom and light neutralino (left) and for well separated gluino and sbottom and heavy neutralino (right). The black curves represent the physical  $b$-jet energy  distributions. The dashed red and blue lines represent the individual  contributions from each step of the decay chains.}
\label{toyspectra}
\end{center}
\end{figure}

We specialize the general strategy outlined above to the case of  pair production of SUSY gluinos and their decay to a $b$ quark along with an on-shell bottom squark which in turn decays into a $b$ quark and a lightest neutralino:
\bea
pp\rightarrow \tilde{g}\tilde{g}\rightarrow bb \tilde{b}\tilde{b} \rightarrow 4b\,2\chi_1^0\,.\label{gluinoTbbbb}
\eea
In the notation of the previous section particle $C$ is the gluino $\tilde{g}$, particle $B$ is the sbottom $\tilde{b}$, the invisible particle $A$ is the lightest neutralino $\chi_{1}^{0}$, and both the final state massless particles $a$ and $b$ are bottom quarks.

We stress that our application to a SUSY case is simply a way to show in detail our mass measurement technique and all its practical aspects. By no means, the applicability of this technique is restricted to the case of SUSY. Furthermore, possible current and future bounds on the existence of the SUSY particles that we discuss leave unchanged our results on the usefulness of a energy peak analysis to extract the mass of new physics particles. 
However, for the sake of completeness we recall that the current limits from the LHC on gluinos and sbottom on the process Eq.~(\ref{gluinoTbbbb}) are around 1.2~TeV for a decay mediated by off-shell sbottom squarks~\cite{CMS-Collaboration:2013bo,CMS-Collaboration:2013pd,ATLAS-CONF-2013-061}. We remark that for our process we assume that the gluino decays via an on-shell squark, which implies that a slightly different bound should apply. In fact, for a generic spectrum that allows a gluino decay via an on-shell sbottom squark, the precise limit will depend on the hypothesized sbottom mass. For instance, for the sbottom mass degenerate to the gluino mass it is likely that the relevant bounds are significantly relaxed due to the softness of some of the final states. In the following we consider concrete examples in which the gluino mass is 1 TeV that is about at the edge of the current limits from the LHC for most choices of the sbottom mass.

As remarked above, we have chosen to apply our technique to the process Eq.~(\ref{gluinoTbbbb}), which corresponds to the emission of an identical SM massless particle at each step of the decay chain. This choice somewhat complicates the subsequent analysis because the energy distributions from the particles emitted at each step of the decay chain will generate its own peak in the inclusive energy distribution of the $b$ quarks. However, we show in the following that our technique is robust enough to deal with this issue.
To exemplify the possible situation that may arise in a realistic situation we pick two choices of the  spectrum of the gluino, sbottom, and neutralino. These are chosen  to illustrate possible issues that arise in the analysis and at the same time  cover the several possible types of spectra that can arise in realistic scenarios such as SUSY. From these considerations we are led to consider two spectra as described below.

\subsection{Spectrum I: $m_{\tilde{g}} \approx m_{\tilde{b}} \gg m_{\chi}$ \label{spectrumone}}
In this case from \eqs{eq:peaka} and~(\ref{eq:peakb}) we expect the two peaks in the energy distribution to be very well separated. We denote the high-energy peak by $E_{HP}$ and the low-energy one by $E_{LP}$. A schematic decomposition of the energy spectrum that we expect from this type of mass spectrum is displayed in the left panel of Figure~\ref{toyspectra}.

We remark that the separation of the two peaks certainly helps to resolve the two peaks individually, but it also poses a couple of challenges. In fact, for this type of spectrum the degeneracy among some of the states makes very likely that at least one emitted $b$-jets is soft.  This poses a problem in that the soft transverse momentum of these $b$-jets may prevent the event from passing the experimental triggers. Furthermore, even when the events are recorded, at low transverse momentum the backgrounds are generically more important than at high transverse momentum. Therefore, it is rather likely that the energy peak arising from very degenerate spectra lies at an energy where the background is too large to observe the peak. In practice, this means that the peak may lie at an energy that is cut away by the transverse momentum requirements that the experiments need to apply to isolate the signal from the backgrounds.
In the following we fit our template to the data available above the thresholds imposed by the cuts required to isolate the signal. The fitted function would describe the entire signal shape, including the part that has been cut away. Therefore, it proves very useful  to have a reliable fitting function to infer the peak position using only the data in the tail.

Another challenge posed by this type of spectrum has to do with the modeling of the signal shape. In fact, a large separation of the two peaks implies that each peak sticks out not only from the background but also from the tail of the other peak present in the energy distribution. As we remarked already, the template for the signal shape found in Ref.~\cite{Agashe:2012mc} is very good over a rather large range of energies around the peak, but it eventually fails to accurately reproduce the shape of the energy distribution when one looks at energies a few times smaller or larger than the peak energy. Therefore, when dealing with the type of energy spectrum that arises from degenerate mass spectra we need to take care of this mis-modeling of the shape of one peak in the region of energies around the other peak.

The degeneracy between gluino and sbottom that characterizes this spectrum, together with the current limits on light new colored states, induces us to consider the gluino and the sbottom much heavier than the neutralino. This means that in any frame the mass of the neutralino is negligible compared to the energy released by the decay of the sbottom. For this reason it is natural to expect that the mass of the neutralino has in general a limited impact on the kinematics of the event. Therefore, even if we have enough relations to invert and determine all three masses by the set of equations in~(\ref{massesfromobservables}), for this spectrum is hard to have a good sensitivity to the neutralino mass. 

 \begin{figure}[t!]
\begin{center}
\includegraphics[width=0.49\linewidth]{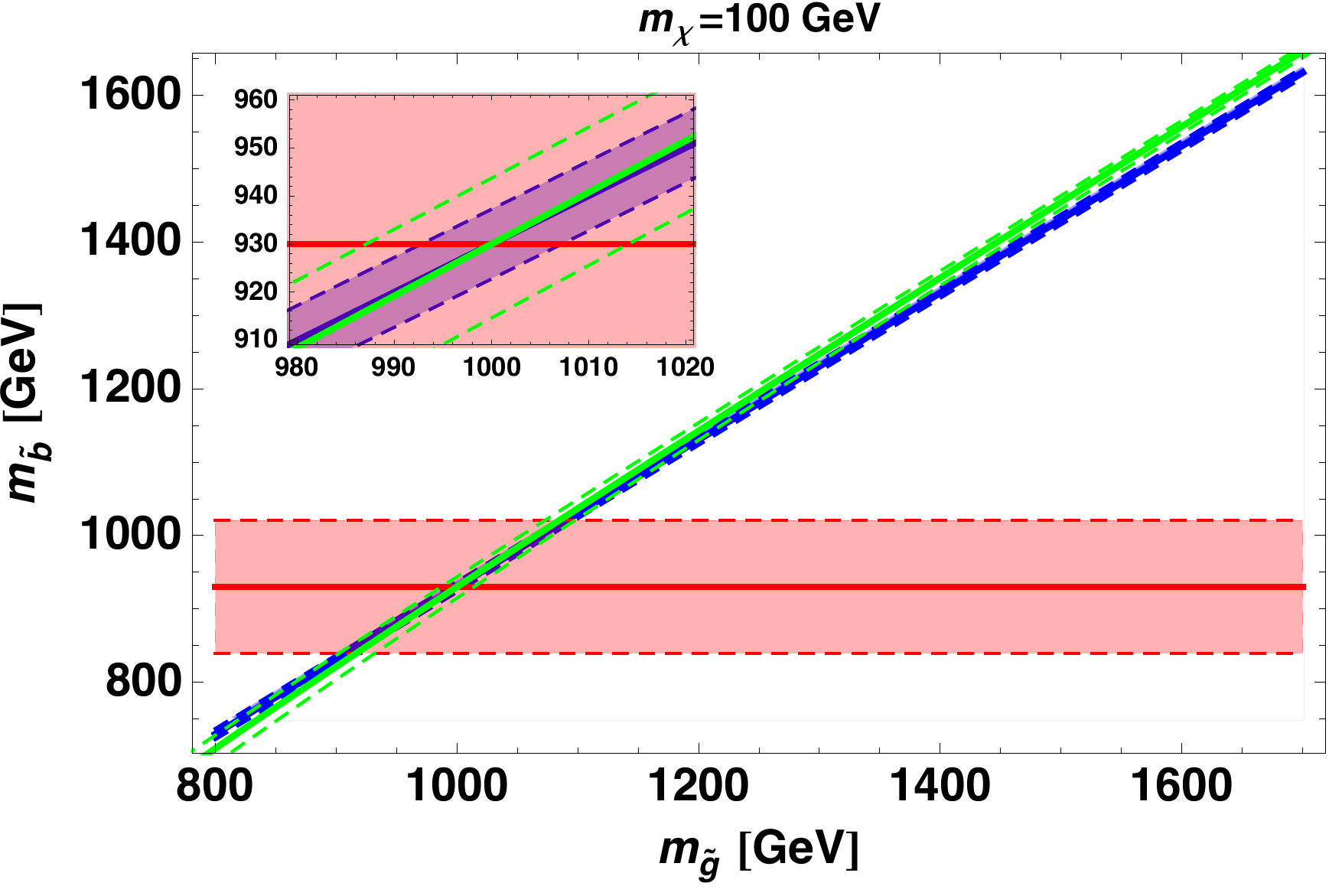}\\
\includegraphics[width=0.49\linewidth]{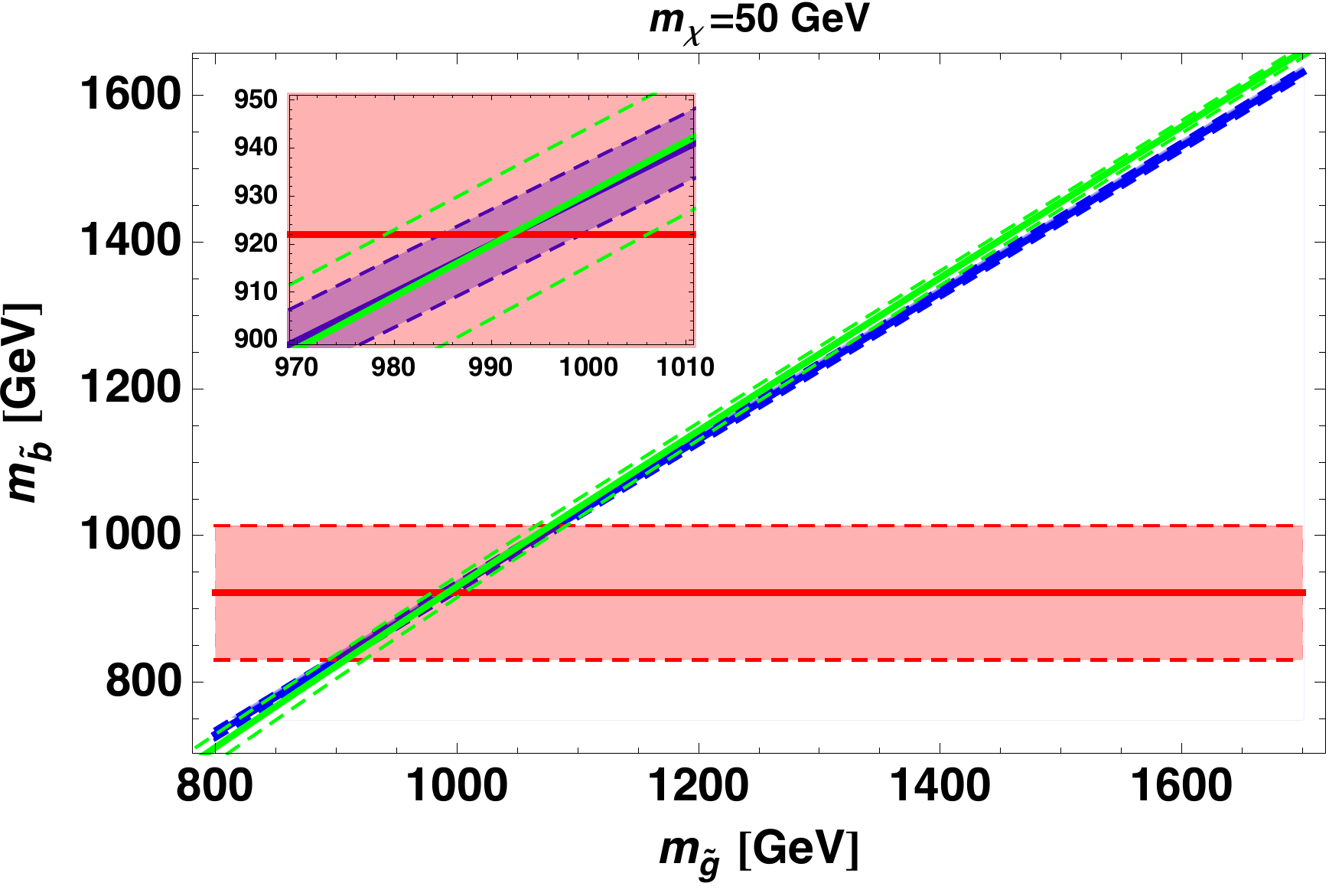}
\includegraphics[width=0.49\linewidth]{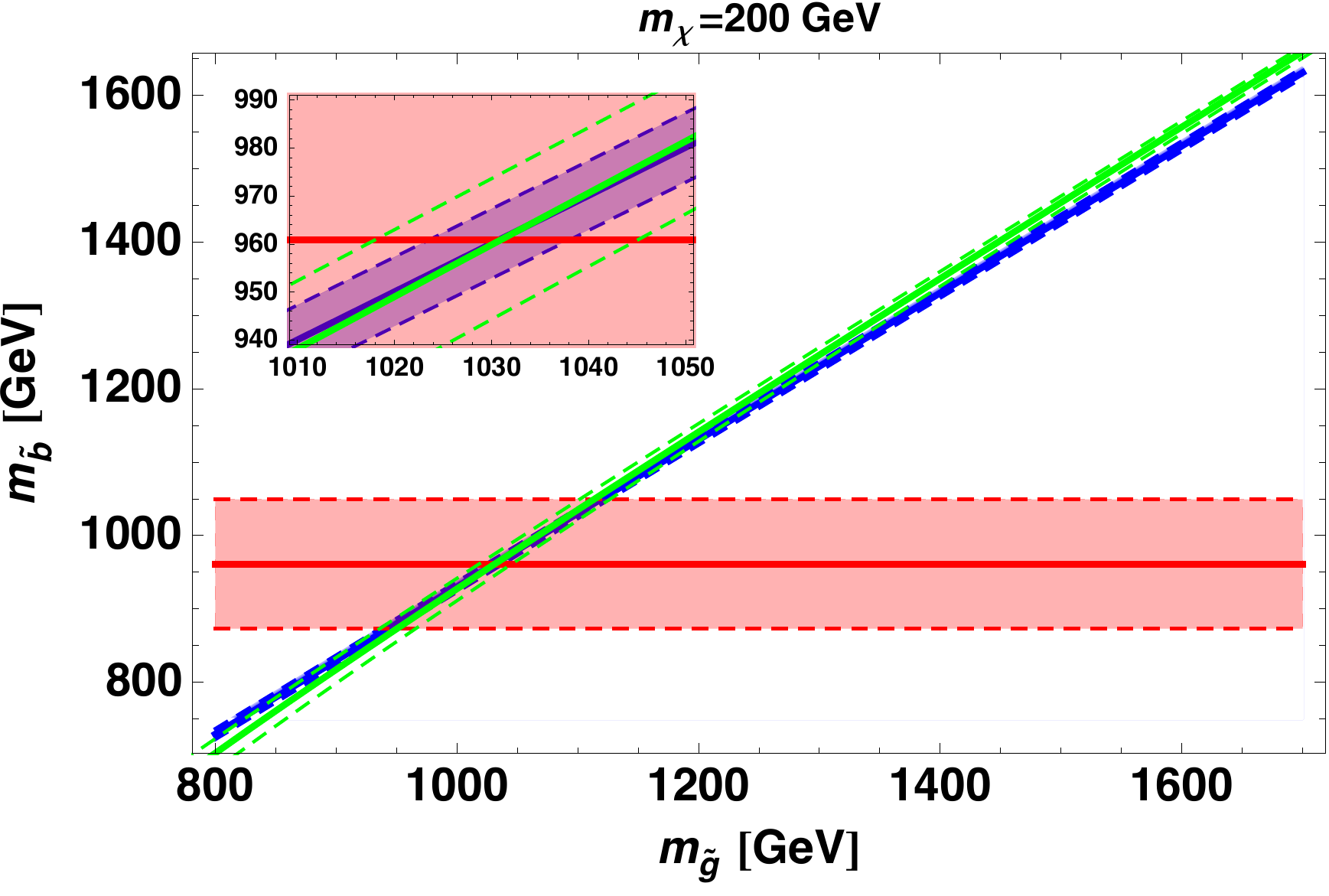}
\caption{Constraints from \eqs{eq:peaka}-(\ref{eq:invedge}) on the plane $m_{\tb},m_{\tg}$ for different choices of $m_{\chi}$.  %
For illustration purposes the measurements of the three observables $E_{LP},E_{HP},m_{bb}$ are assumed to be at the theory values from the \eqs{eq:peaka}-(\ref{eq:invedge}) and masses as in \eq{massesone} (solid lines). A 10\% uncertainty in the observables is assumed, which is reflected in the fact that each constraint is satisfied on a band in the plane $m_{\tg},m_{\tb}$ delimited by dashed lines. For the observables that are sensitive to it, $m_{bb}$ and $E_{HP}$, the neutralino mass $m_{\chi}$ has been fixed as indicated in each panel. The inset shows a close-up of the region around the point where the three constraints come close to each other. The red band represents the constraint from $E_{HP}$, the blue represents the one from $E_{LP}$, and the green the one from $m_{bb}$. {The area where the three constraints overlap identifies the measured $m_{\tg}$ and $m_{\tb}$.}
\label{contraintsplanecaseone}  
}
\end{center}
\end{figure}

{In order to gauge the achievable sensitivity to the neutralino mass it is useful to go through an exercise. For this spectrum it is useful to expand eq.(\ref{eq:peaka}) around $m_{\tilde{g}}\simeq m_{\tilde{b}}$, which gives 
\begin{equation}
E^{*}=\left( m_{\tilde{g}} -m_{\tilde{b}}\right)+\frac{\left(m_{\tilde{g}}-m_{\tilde{b}}\right)^2}{
   2 m_{\tilde{g}}}\,.\label{EstarSmallMassDifference}
\end{equation}
 In eq.(\ref{eq:invedge})  one can solve for $m_{\tilde{g}}$ and take the dominant piece for $m_{\tilde{b}}\gg m_{bb},m_{\chi}$, so that the solution reads 
\beq
m_{\tilde{g}}\simeq m_{\tilde{b}}+\frac{m_{bb}^{2}}{2
   m_{\tilde{b}}} +\frac{
   4m_{\chi}^2 m_{bb}^{2}- m_{bb}^{4}}{8m_{\tilde{b}}^3}\,. \label{largembMbbconstraint}
\eeq

From the two above equations is clear that the constraints on the masses from the two observables $E^{*}$ and $m_{bb}$ are highly correlated. In fact both the equations (\ref{EstarSmallMassDifference}) and (\ref{largembMbbconstraint}) can be casted in the form 
\beq
m_{\tg} \simeq m_{\tb} + \epsilon_{1,2}\,,  
\eeq
where both $\epsilon_{1}$ and $\epsilon_{2}$ are much smaller than $m_{\tg}$ and $m_{\tb}$.
Therefore a poor mass determination should be anticipated because of the large degree of parallelism of the two constraints.
To confirm this analytical finding in Figure~\ref{contraintsplanecaseone} we show the constraints from \eqs{eq:peaka}-(\ref{eq:invedge}) on the plane $m_{\tb},m_{\tg}$ for different choices of $m_{\chi}$. As it can be appreciated from the figure the constraints from the energy peak of the $\tg\to\tb b$ decay (blue) and the $m_{bb}$ (green) are almost parallel even for quite large variations of the assumed mass of the neutralino. 
Furthermore from the picture we can see that variations of order 100\% on the assumed mass of the neutralino do not affect significantly the relative positions of the three lines from the three constraints. In fact for $m_{\chi}=200\gev$ as well as $m_{\chi}=50\gev$ all the three lines cross at one point, as it should for the constraints evaluated at the correct neutralino mass. This means that the set of observables that we used to extract the masses for this spectrum is basically insensitive to the neutralino mass, as the lines continue to cross almost perfectly even when the neutralino mass is 100\% different from the correct value~\footnote{{We remark that the unfavorable parallel nature of two constraints might be avoided picking another observables in place of $m_{bb}$. For instance the third observable might be another energy peak: the energy peak of the compound system made of the two $b$-jets for which an extension of the theory results in Section~\ref{theory} is available~\cite{massivetemplate,threebody}. Picking an energy peak as third observable the masses would, very nicely, be reconstructed  using just energy peaks.}}.}

\begin{figure}[t!]
\begin{center}
\includegraphics[width=0.49\linewidth]{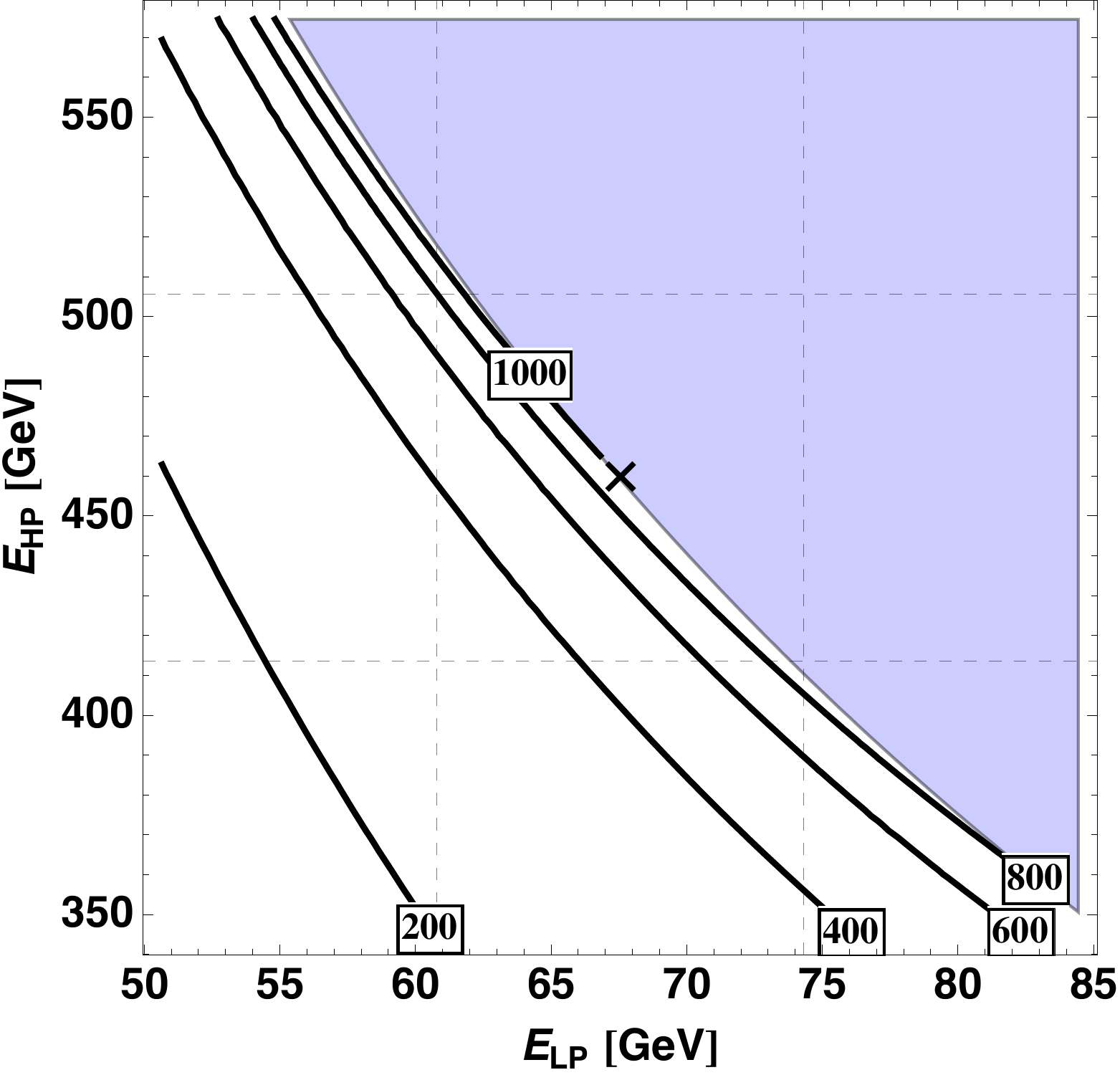}
\caption{Mass of the gluino as a function of the measured energy peaks for fixed $m_{bb}$. The black cross denotes the theory value of the energy peak following from the true masses in \eq{massesone}. The dashed lines delimit the 10\% variations from the theory value.
\label{sensitivitytoElpEhp}  
}
\end{center}
\end{figure}

{Another related issue that arises when the gluino and the sbottom are degenerate has to do with the physical viability of energy peaks values in the vicinity of the correct one.  In fact if we take \eq{massesfromobservables} we can see that the masses of the sbottom and the gluino ($m_{B}$ and $m_{C}$ in \eq{massesfromobservables}) suffer an instability when $m_{\tb}\to m_{\tg}$ . In that case the numerator and the denominator both vanish as a consequence of both $m_{bb}$ and $\Ebpeak$ vanishing. As a  result the values of  $m_{\tb}$ and $m_{\tg}$ computed from the energy peaks and $m_{bb}$ are extremely sensitive to the precise value of the fitted energy peaks. To better appreciate this sensitivity we show in Figure~\ref{sensitivitytoElpEhp} the isolines of $m_{\tg}$ in the plane $E_{LP},E_{HP}$. The figure also shows the region that should be cut out from the plane because the inequality \eq{inequalities} necessary to obtain physical masses is not satisfied. A similar sensitivity appears for the sbottom mass but we do not show the related plot that just looks the same as the one for the gluino.

This sensitivity to the precise measured value of the energy peaks exposes our method for the mass measurement to possible large uncertainties on the masses even in presence of small error on the peak determination.  It should be noted that this issue is in part related to do fact that we do not vary $m_{bb}$ when we consider the sensitivity of $m_{\tg}$ to the energy peaks position. In reality the experimentally determined position of $m_{bb}$ could be shifted from the theory value. Part of this shift is due to physics reasons that are in a certain degree of  correlation with  the mismeasurement of the energy peaks. However we expect that the sharpness of the $m_{bb}$ edge will allow a determination of $m_{bb}$ significantly more precise than that of the energy peaks, hence justifying our simplified treatment in Figure~\ref{sensitivitytoElpEhp}.}

{The above results about the sensitivity to small errors in the energy peak determination and the parallelism of the constraints on the masses from different observables motivates us to not attempt to use \eq{massesfromobservables} for the mass spectrum with almost degenerate gluino and sbottom. We proceed by simplifying the analysis and putting aside for the time being the mass determination of the neutralino, on which we return later. Therefore for this spectrum we attempt a measurement of the gluino and sbottom masses under the simple assumption that the neutralino is just massless. This assumption, as a flip side of the the previous observations on the crossing of the constraints, has limited, but in general not negligible, impact on the mass determination for $m_{\tg}$ and $m_{\tb}$.
As can be checked from \eqs{eq:peaka}-(\ref{eq:invedge}) and from Figure~\ref{contraintsplanecaseone}, taking a massless neutralino at the bottom of the spectrum  in general results in an underestimation of $m_{\tg}$ and $m_{\tb}$, which can be easily several percent off from the true values. For the time being we do not seek  a percent precision mass determination, which is certainly premature for new physics and even more so for our new method. Therefore we do not comment further on the possible underestimation of the gluino and sbottom masses.
}
  
{Taking a massless neutralino 
the general formulae \eqs{massesfromobservables} get reduced to the simpler relations. If one wishes to use the two quantities that have the best chance to be more precisely determined from the data, i.e. $m_{bb}$ and $E_{HP}$, then the relevant equations for the masses of gluino and sbottom are
\bea
m_{\tilde{b}}=2E_{HP},\;\;\;m_{\tilde{g}}=\sqrt{m_{bb}^2+4E_{HP}^2}\,. \label{simplifiedcaseone}
\eea
We remark that the masses measured from these relations, compared to \eqs{massesfromobservables}, are not highly sensitive to  small uncertainties in the determination of the energy peak $E_{HP}$. Indeed when compared to \eqs{massesfromobservables} they involve much lower powers of the observable quantities and therefore the error propagation benefits as well from this approximation.
Alternatively one could express the masses only as functions of the energy peaks  $E_{HP}$ and $E_{LP}$,  the latter having  some more experimental obstacles to face if one aims for a precise measurement. Despite the experimental challenges posed by the determination of $E_{LP}$, for instance by acceptance cuts, the possibility of using just the energy peaks is anyhow noteworthy because then the masses can be reconstructed  relying only on the novel observables  that we consider in this paper. The inversion relations in this case are as follows:
\beq
m_{\tilde{b}}=2E_{HP},\;\;\;m_{\tilde{g}}=E_{LP}+\sqrt{E_{LP}^2+4E_{HP}^2}\,. \label{simplifiedcaseoneenergy}
\eeq}

\begin{figure}[t!]
\begin{center}
\includegraphics[width=0.49\linewidth]{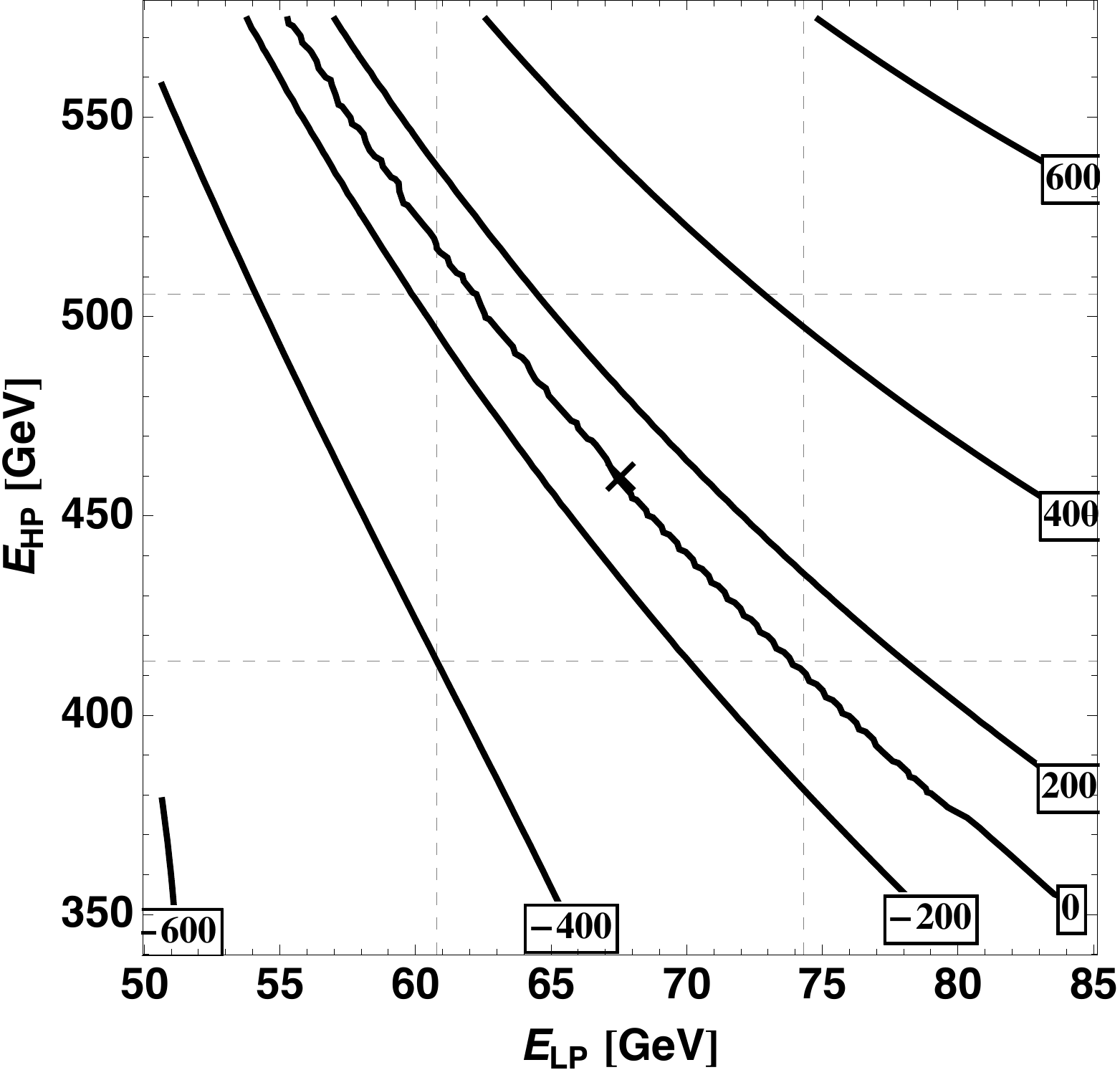}
\caption{Signed-Mass of the neutralino, i.e. $\textrm{sign}(m_{\chi}^{2})\sqrt{|m_{\chi}^{2}|}$, as a function of the measured energy peaks for fixed $m_{bb}$ as found by \eq{mchiexactfunctionofpeaks}. The black cross denotes the theory value of the energy peak following from the true masses in \eq{massesone}. The dashed lines delimit the 10\% variations from the theory value.
\label{mChivsElpEhp}  
}
\end{center}
\end{figure}

{Having simplified the problem, the knowledge about $m_{\tg}$ and $m_{\tb}$ can be used to attempt to recover some information about $m_{\chi}$.}
{Inverting \eq{eq:invedge} we obtain that 
\beq
m_{\chi }^2= \frac{m_{\tilde{b}}^2
   \left(m_{\tilde{b}}^2-m_{\tilde{g}}^2+m_{bb}^2\right)}{m_{\tilde{b}}^2-m_{\tilde{
   g}}^2}
\eeq
and using the estimates for the gluino and sbottom mass from \eq{simplifiedcaseoneenergy} we can express the neutralino mass as
\beq
m_{\chi }^2=\frac{4 E_{{HP}}{}^2 \left(m_{{bb}}^2-\left(\sqrt{4
   E_{{HP}}{}^2+E_{{LP}}{}^2}+E_{{LP}}\right){}^2+4
   E_{{HP}}{}^2\right)}{4 E_{{HP}}{}^2-\left(\sqrt{4
   E_{{HP}}{}^2+E_{{LP}}{}^2}+E_{{LP}}\right){}^2}\,.
\label{mchiexactfunctionofpeaks}
\eeq
This expression has the notable property to be significantly more stable than the corresponding one in \eq{massesfromobservables} when small variations of the measured energy peaks are considered. In Figure~\ref{mChivsElpEhp} we show the dependence of the reconstructed $m_{\chi}$ in the plane $E_{LP}$, $E_{HP}$. From the figure we see that a moderate dependence on the precise value of the energy peaks is still present. However the figure shows that with a knowledge at 10\% of the energy peaks one should be able to exclude a neutralino mass around 400~GeV. We find remarkable that a mass scale estimate, although quite rough, can be attained.}
   
\subsection{Spectrum II: $m_{\tilde{g}} \gtrsim m_{\tilde{b}} \gtrsim m_{\chi}$\label{spectrumtwo}}
In this case the expected mass difference induces a large average energy release at each step of the decay chain, and typically all the $b$-jets have comparable energies in the laboratory frame. Therefore, we expect that the energy distributions of the $b$-jets arising from the gluino decay and that arising from the sbottom decay largely overlap. A schematic decomposition of the energy spectrum that we expect from this type of mass spectrum is displayed in the right panel of Figure~\ref{toyspectra}.
 
\begin{figure}[t!]
\begin{center}
\includegraphics[width=0.49\linewidth]{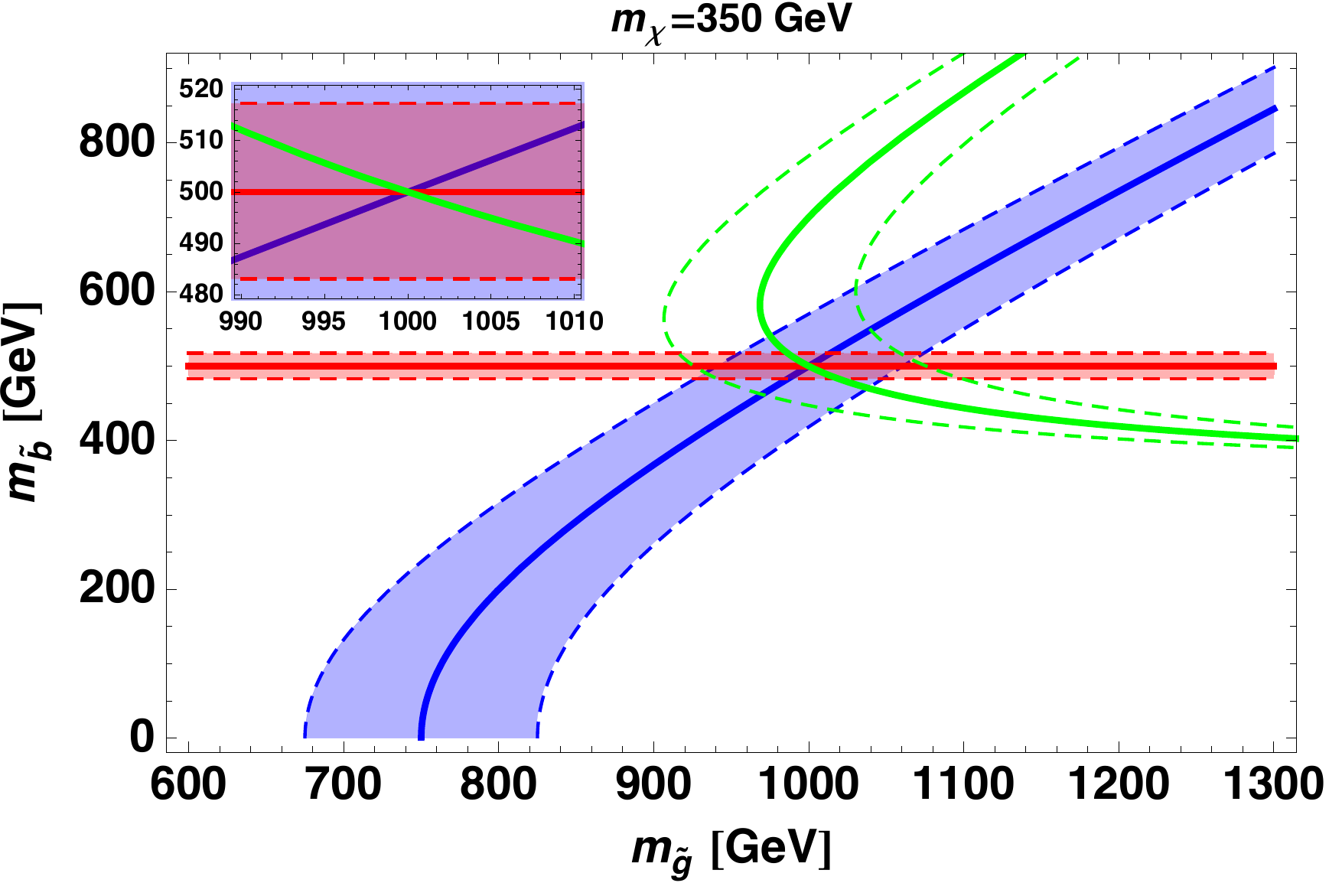}\\
\includegraphics[width=0.49\linewidth]{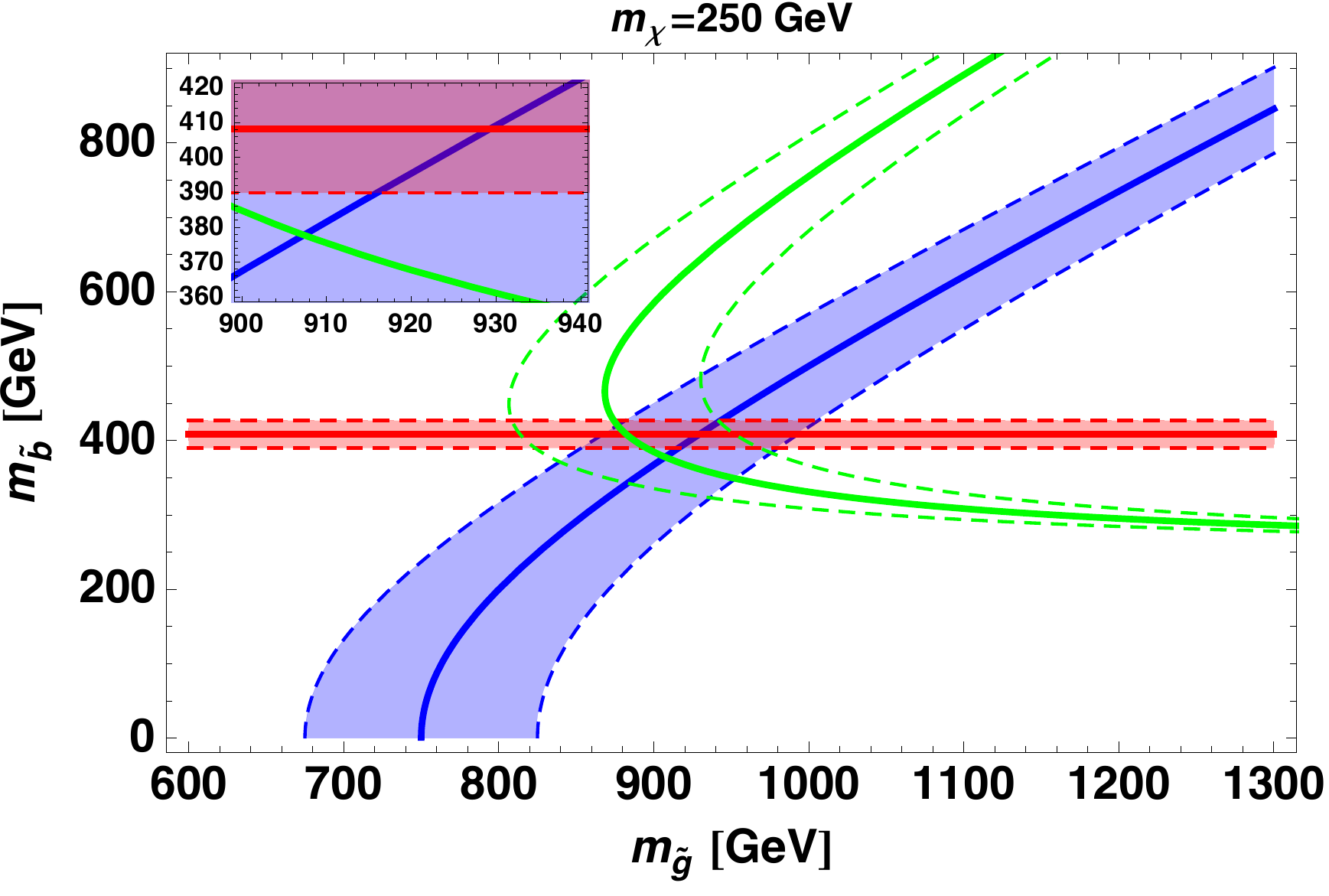}
\includegraphics[width=0.49\linewidth]{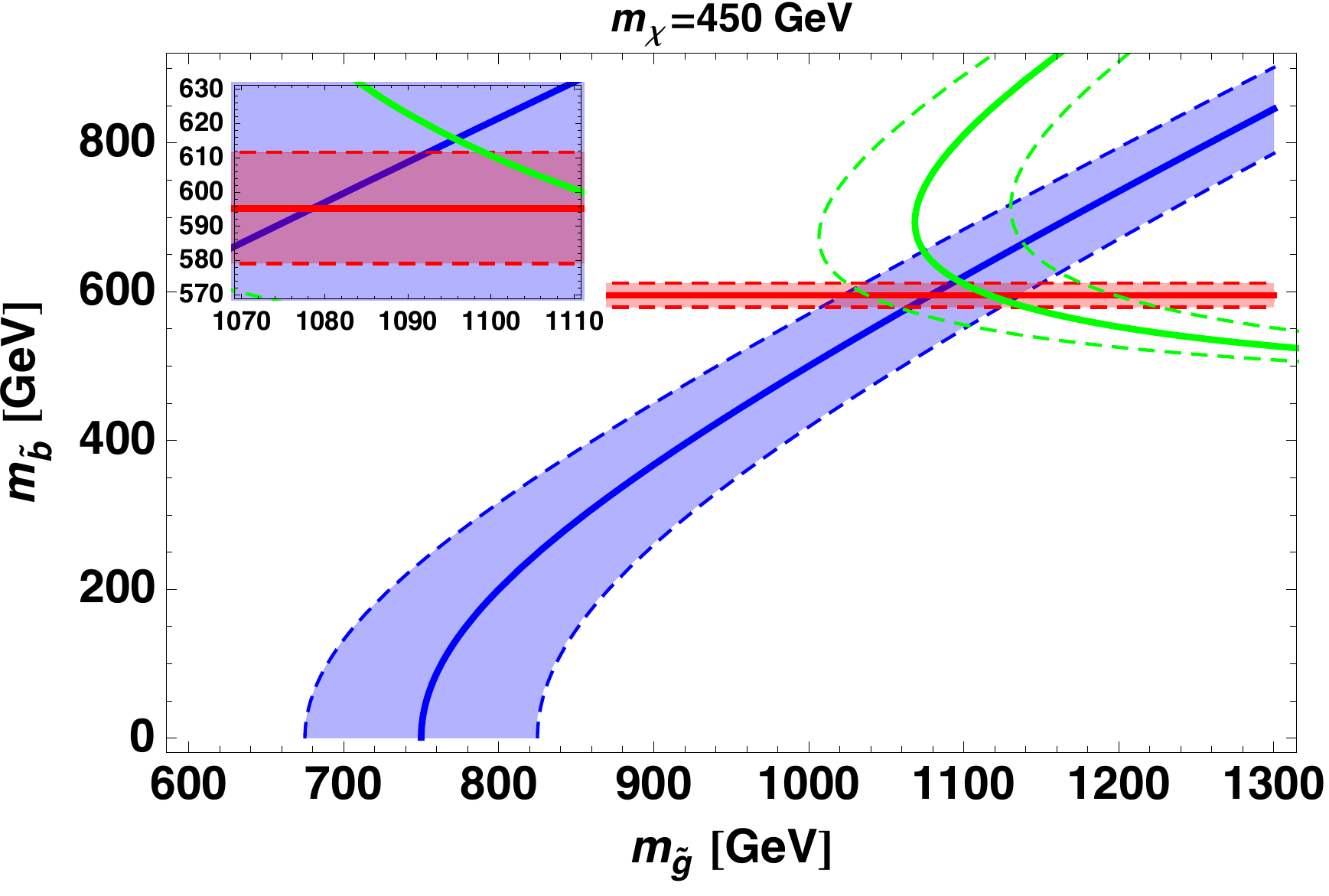}
\caption{Constraints from \eqs{eq:peaka}-(\ref{eq:invedge}) on the plane $m_{\tb},m_{\tg}$ for different choices of $m_{\chi}$. %
For illustration purposes the measurements of the three observables $E_{LP},E_{HP},m_{bb}$ are assumed to be at the theory values from the \eqs{eq:peaka}-(\ref{eq:invedge}) and masses as in \eq{massestwo} (solid lines). A 10\% uncertainty in the observables is assumed, which is reflected in the fact that each constraint is satisfied on a band in the plane $m_{\tg},m_{\tb}$ delimited by dashed lines. For the observables that are sensitive to it, $m_{bb}$ and $E_{HP}$, the neutralino mass $m_{\chi}$ has been fixed as indicated in each panel. The inset shows a close-up of the region around the point where the three constraints come close to each other. The red band represents the constraint from $E_{HP}$, the blue represents the one from $E_{LP}$, and the green the one from $m_{bb}$. {The area where the three constraints overlap identifies the measured $m_{\tg}$ and $m_{\tb}$.}
\label{contraintsplanecasetwo}  
}
\end{center}
\end{figure}

The challenge for this spectrum comes from the fact that each peak is typically broad, due to the non negligible boost of each mother particle and the large energy releases at each step of the decay chain. In general, the observed energy spectrum for this kind of well separated mass spectra has a single bump, which results from overlaying of the two peaks coming from the two steps of the decay chain. Therefore, it is very hard to guess the double peak structure by simply looking at the energy spectrum. On the contrary, using reliable fitting functions such as that in \eq{eq:fitter} we are able to resolve the two peaks and extract the masses of the particles involved in the decay.
Given that at each step of the decay the masses of new particles are all comparable, we expect that any kinematic analysis with the relevant final state has chances to have the sensitivity to all three masses involved. 
{This intuition is confirmed by the study of the constraints \eqs{eq:peaka}-(\ref{eq:invedge}) in the plane $m_{\tg},m_{\tb}$. In Figure~\ref{contraintsplanecasetwo} we show these constraints for different choices of $m_{\chi}$ assuming the energy peaks and the $m_{bb}$ edge for a spectrum $m_{\tilde{g}} \gtrsim m_{\tilde{b}} \gtrsim m_{\chi}$ given in \eq{massestwo}. We can see from the figure that the three constraints cross each other at an angle, which implies that for this choice of spectrum they are largely independent constraints. Not surprisingly, varying the value of the hypothetical $m_{\chi}$ around the correct value $m_{\chi}=350\gev$ the three constraints cross at two points that are noticeably separated. This indicates that a certain sensitivity to $m_{\chi}$ can be attained with these three observables.} For this reason, unlike for what we did the previous Section, here we make use of the exact relations given by  \eqs{massesfromobservables}.

\subsection{Events simulation and selections\label{simulation}}

In order to test our strategy and quantify the accuracy that can be reached by a mass measurement that uses energy peaks as input, we apply our techinque to simulated events of gluino production at the LHC including the relevant background processes.

Our signal process is defined in \eq{gluinoTbbbb} and we fix the two following mass spectra for the two classes discussed before:
\begin{description} 
\item[{\bf \quad Spectrum I}] 
\beq m_{\tilde{g}}=1 \TeV,  m_{\tilde{b}}=930\GeV,\textrm{ and }m_{\chi_1^0}=100\GeV\,,\label{massesone}\eeq 
\noindent
from which we expect two peaks at $E_{LP}=68\gev$ and $E_{HP}=460\gev$;
\item[{\bf \quad Spectrum II}] 
\beq m_{\tilde{g}}=1 \TeV,  m_{\tilde{b}}=500\GeV,\textrm{ and }m_{\chi_1^0}=350\GeV\,,\label{massestwo}\eeq 
\noindent
from which we expect two peaks at $E_{LP}=127\gev$ and $E_{HP}=375\gev$\,.
\end{description}
In each signal event there are four $b$ quarks, which give rise to jets, and two invisible neutralinos that result in large missing transverse energy.  In what follows we consider as signal the subset of events coming from gluino production that result in the signature $$4b+\misse\,,$$
where we have required 4 jets to be reconstructed and tagged as $b$-jets. We treat the issue of $b$-tagging at a simplified level, which is sufficiently accurate for our purposes, and we  assume a tagging efficiency constant in the $\eta-p_{T}$ plane and equal to 0.66~\cite{CMS-Collaboration:2012uq,ATLAS-CONF-2012-043,ATLAS-CONF-2012-097}. 

The major background processes are~\footnote{In principle, also the multi-jet production from pure QCD $pp\to 4b$ would constitute a background, whose $\misse$ arises from mismeasurements of the jet energy and direction (we thank Lian-Tao Wang for pointing out this background). This background is particularly difficult to estimate since it is completely due to detector effects. However, in the following we conceive cuts \eq{metcut} and \eq{dphicut} that have great rejection power against this type of instrumental background. We have studied event samples where these detector effects are emulated using $\mathtt{Delphes 1.9}$~\cite{Ovyn:2009ys}, and found that this type of background is sub-dominant with respect to the ones considered in the rest of the paper.} $$pp\to Z\,4b\to \nu\bar{\nu}\,4b \textrm{\quad and \quad} pp\to t\bar{t}\,b\bar{b}\,.$$ The former is irreducible whereas  the latter process has  a different partonic final state. Despite the different partonic final state the process $pp\to t\bar{t}\,b\bar{b}$  becomes a background for our final state when some partons are ``lost''. This means that the visible products from the decay of the two $W$ bosons are not seen by the detector either because they did not pass the acceptance due to their low $p_T$ or large $\eta$ or both, or because they were not sufficiently isolated from the rest of the hard particles in the event so that they had been merged with other objects in the event. To take into account this kind of effects we define as a missed parton any object with any of the following properties:
\begin{itemize} \itemsep0.5pt
\item for jets, $p_{T,j}<30$~GeV or $|\eta_{j}|>5$,
\item for leptons, $p_{T,l}<10$~GeV or $|\eta_{l}|>3$.
\end{itemize}
To model the part of the backgrounds that come from non-isolated objects being merged in the detector reconstruction we use the following criteria:
\begin{itemize}
\item for merging jets, $\Delta R_{j_1j_2}<0.4$ with $j_1j_2$ denoting any jet pairs including $b$-jets,
\item for merging leptons, $\Delta R_{jl}<0.3$ with $j$ and $l$ denoting a jet and a lepton, respectively.
\end{itemize}
The $t\bar{t}b\bar{b}$ process is a pure strong interaction process and in principle could be the dominant background. Due to acceptance and isolation requirements just described most of the background events originate from the fully leptonic and the semi-leptonic decay channels of top quark pairs because fewer partons need to be lost compared to the  fully hadronic top decay channel. 

In general, we expect that the production of new heavy particles gives rise to jets with larger transverse momentum than those arising from standard model events. This is in general a good way to roughly discriminate between new physics and SM events. However, it must be remarked that for our purpose of measuring masses by searching for peaks in the energy distribution we have to be careful not to distort the energy distribution, which would spoil our method. Therefore, in what follows we avoid pushing too hard in requiring hard single objects in the final state to isolate the signal from background events.
For identification purposes we require 
\beq
p_{T,b}>30 \GeV, |\eta_{b}|<5, \Delta R_{bb}>0.4\,\label{preselection} 
\eeq
in all the events that we use for our analysis. Furthermore, to reject efficiently background events while retaining a large fraction of the signal event we exploit the tendency of the signal to give large missing transverse energy, roughly set by some combination of the new particles masses. For the backgrounds the missing transverse energy is set by whichever is the largest between the mass of the $Z$ ($t$ quark) and the total hardness of the event. Therefore, signal isolation can be achieved by requiring a large $\misse$. 
It is useful to notice that in general the missing transverse energy is the result of the invisible particles recoiling against visible ones. For this reason when one requires a large $\misse$, automatically the hardness of the visible particles increases as well. This large $\misse$ requirement could in principle hamper our mass measurement strategy by inducing a too large bias in the $b$-jet energy spectrum. Fortunately, for the signal it is quite likely to have multiple hard objects that are collectively giving the large recoil to the invisible particles. Therefore, we expect only a modest bias in the energy spectrum coming from the $\misse$ selection.  We find that a rather strong reduction of the background, still without significant distortion in the $b$-jet energy spectrum, can be attained by requiring
\beq
\misse >150\GeV\label{metcut}\,.
\eeq
Additionally, we require each $b$-jet to be sufficiently distant in azimuthal angle from the direction pointed by the $\misse$ vector. By doing this we make sure that the measured $\misse$ is not arising from mismeasured jet(s). For our study we require
\beq
\Delta\phi(\misse,j)>0.2\,\label{dphicut}
\eeq for all the $b$-jets.

To evaluate the cross-section and the energy distributions for signal and background we produced simulated samples of $pp$ collisions at the 14 TeV LHC using $\mathtt{MadGraph5}$~\cite{Alwall:2011uj}. The structure of the proton is parametrized by the parton distribution functions (PDFs)  $\mathtt{CTEQ6L1}$~\cite{Pumplin:2002vw} evaluated with a renormalization and factorization scale varied depending on the kinematics of each event according to the default of  $\mathtt{MadGraph5}$. 
\begin{table}[t]
\centering
\begin{tabular}{|c||c|c|c|c|}
\hline
 & Spectrum I & Spectrum II & $Z\,4b$ & $t\bar{t}b\bar{b}$ \\
\hline \hline
Cross-section [fb] & 94.1 & 108.7 & 1.15 & 0.41 \\
\hline
\end{tabular}
\caption{The expected cross sections for signal and background events after imposing the cuts \eqs{preselection}-(\ref{dphicut}) plus the isolation and identification criteria described in the text. The efficiency of $b$ tagging is not taken into account in this table. The effect of $b$ tagging efficiency should be the same on the four columns since the number of $b$ quarks in each process is the same and $b$-tagging efficiency is assumed universal in the $p_T$-$\eta$ plane.}\label{tab:compare}
\end{table}  
The resulting total cross-section are reported in Table~\ref{tab:compare}, which clearly shows that the background from $t\bar{t}b\bar{b}$ is sub-dominant compared with $Z+4b$, although just by a factor a few. In the following we proceed to a simplified analysis in which we ignore the $t\bar{t}b\bar{b}$ background, and retain only $Z+4b$ as the dominant background. In principle the $t\bar{t}b\bar{b}$ can be added to the analysis, hence changing the details of our study, but without any major impact in the results. 
From the table we also see that the signal-over-background ($S/B$) for both types of mass spectrum is quite large. 
This of course renders our job of extracting the masses of the new particles significantly easier. However, when presenting results in Section~\ref{Results}, we will comment about possible less favorable $S/B$.

\subsection{Energy peak fitting strategy\label{fittingstrategy}}
As mentioned in the Introduction, we extract the peaks of the energy distribution using  the fitting function given in Eq.~(\ref{eq:fitter}). For each different type of mass spectrum the resulting energy spectrum poses different challenges described in Sections~\ref{spectrumone} and \ref{spectrumtwo} which we address as explained in the following. In all cases we perform a simultaneous fit to the data with a fitting template that includes contributions from the background and from the $b$-jet energy distributions expected by each of the two steps in the decay chain. 
The background is modeled by a function
\bea
f_{\mathrm{BG}}(E)=N_{b}\exp\left(-b \cdot \sqrt{E} \right)\label{eq:bgfittergen}\,,
\eea
where $b$ is a fit parameter that determines the shape of the functions, and $N_{b}$ is a fit parameter responsible for the total number of events described by the function (at fixed $b$).
This function has been tested on samples of pure background and describes very well the energy distribution {\it after} the selections described in \eqs{preselection}-(\ref{dphicut}). In fact, fitting this function to simulated data over the several different ranges of energy that we use in the following to extract the peaks from the signal, we found that this function captures the background shape well enough and typically yields a reduced $\chi^{2}\simeq1$ when compared with the simulated data. Similar types of exponential functions are frequently used in simultaneous fits of signal and background to data~\cite{Bayatian:942733,ATLAS:1999vwa}.
In our numerical study for the two mass spectra introduced above we assume that $N_{b}$ and $b$ have been determined using data-driven methods, for instance, the ABCD method~\cite{Bayatian:942733,ATLAS:1999vwa} or similar ones that allow to fix the properties of the background shapes inferring them from control regions where reliable Monte Carlo predictions are available and there is little signal contamination. 
In this paper, we do not address the issue of the optimal definition of signal and control regions for the data-driven estimate of the background in the signal region. In fact, this type of study belongs more properly to the domain of the experimental collaborations as the details of it will depend quite significantly on the specific experimental conditions. As a substitute for the data-driven prediction, in our study we 
use a leading order Monte Carlo simulation in order to fix the background shape and normalization parameters.
We denote the quantities determined from the Monte Carlo simulation of the background by adding a ``bar'' on each symbol, and thus the fixed background function that we use in the following is given by 
\bea
\bar{f}_{\mathrm{BG}}(E)=\bar{N}_{b}\exp\left(-\bar{b}\cdot \sqrt{E} \right)\label{eq:bgfitter}\,.
\eea
We stress that our background shape from the Monte Carlo is only a lay figure that allows us to account for some of the effects that arise from the presence of the background in the data used to extract the energy peaks. We firmly insist on the fact that in a realistic application of our mass measurement strategy,  
%
%
the background shape should be obtained from the data, which would 
%
%
better 
%
%
account for 
%
%
any effects
%
%
of mismeasurement and acceptance 
%
%
that are poorly described by simulations.

\subsubsection{  Fitting of the energy spectrum for the mass spectrum I}
For the mass spectra where the gluino and the sbottom masses are nearly degenerate and the neutralino is light, we expect a $b$-jet energy spectrum similar to the one sketched in the left panel of Figure~\ref{toyspectra}. 
As can be seen therein, the two peaks in this case are well-separated and we found that it is indeed possible to fit each peak separately. To do this we consider data in a range of energy where one of the two peaks dominates and the other is largely sub-dominant. Then we proceed to fit the data using a template function \eq{eq:bgfitter} to account for the background, plus a peak template of the type \eq{eq:fitter} and a template for the modeling of the tail of the other peak, which we describe in the following. The necessity to model the tail of the sub-dominant peak arises as this tail effectively constitutes a pollution to the extraction of the value of the peak that dominates in this range of energy. We repeat a similar fit for a different energy range where the role of the two peaks is exchanged, i.e. where the previously sub-dominant peak is now the dominant component of the energy spectrum and vice versa.
More in detail, our complete template used to fit  the data around the high energy peak is 
\bea
f_{\mathrm{HP}}(E)+f_{\text{LP}}^{\text{eff}}(E)+\bar{f}_{\mathrm{BG}}(E)\label{templateHP}\,,\label{HPtotaltemplate}
\eea
where
\bea
f_{\textrm{HP}}(E)&=&N_{HP}\exp\left(-\frac{w_{HP}}{2}\left(\frac{E}{E_{HP}}+\frac{E_{HP}}{E} \right) \right)\,,
\eea
which is a template of the type  \eq{eq:fitter} for the peak region,
\bea
f_{\textrm{LP}}^{\textrm{eff}}(E)&=&N_{t}\exp\left(-t E\right)\,,
\eea
which an effective parametrization for the tail of the low-energy peak,
and $\bar{f}_{\mathrm{BG}}$ is given in Eq.~(\ref{eq:bgfitter}). Here $t$ is the fit parameter that affects the shape of the template used to model the tail of the sub-dominant peak, and $N_{t}$ is a parameter that, for fixed $t$, describes the number of events from the tail of the sub-dominant peak.  The parameters that describe the dominant peak are $w_{HP}$, which defines the width of the peak, $E_{HP}$, which defines the position of the sought peak, and finally $N_{HP}$, which sets the total number of events described by the peak. The fit of the template \eq{templateHP} to the data will return a best-fit value for each of the parameters with its own uncertainty due to the fluctuations in the data. From this output of the fit, we use  the best-fit value of $E_{HP}$ and its error as an input for \eq{simplifiedcaseoneenergy}  to compute the masses of the heavy particles and the corresponding uncertainty.

For the low energy peak we pursue a similar approach although the peculiarities of the specific case require us to slightly change our strategy. As discussed in section~\ref{spectrumone} the cuts necessary to isolate the signal from the background tend to modify the low energy part of the $b$-jet energy distribution and in some cases they may even cut away the entire low-energy peak region. The selections \eqs{preselection}-(\ref{dphicut}) that we have used to isolate the signal are sufficiently mild that we still observe a low-energy peak. However, we want to demonstrate that our energy peak strategy can be used even for less favorable cases where the peak cannot be seen at all in the data  as a consequence of the cuts. For this reason in our fit we consider only off-peak data points with energies above the low-energy peak visible in the data. 
Since we want to perform an off-peak analysis of the data to infer the low energy peak position, we need to analyze a rather large range of energies such as to have a substantial number of events in the fit. Thus our choice to perform an off-peak analysis requires to treat with care the contamination from the tail of the high-energy peak, which becomes more important as one widens the range of energies in the data.
As we did for the high-energy peak, in order to deal with this issue we introduce in our fit a function that captures the contribution of this tail in the region around the low-energy peak. The overall template that we use to fit the low-energy data is
\bea
f_{\mathrm{LP}}(E)+f_{\mathrm{HP}}^{\mathrm{eff}}(E)+\bar{f}_{\mathrm{BG}}(E)\,,\label{LPtotaltemplate}
\eea
where
\bea
f_{\mathrm{LP}}(E)\hspace{-0.2cm}&=&\hspace{-0.2cm}N_{LP}\exp\left(-\frac{w_{LP}}{2}\left(\frac{E}{E_{LP}}+\frac{E_{LP}}{E} \right) \right)\,, \label{templateLP} 
\eea
which is essentially the same type of function used to fit the high-energy peak, 
\bea
f_{\mathrm{HP}}^{\mathrm{eff}}(E) \hspace{-0.2cm}&=&\hspace{-0.2cm} N_{t}\exp\left( -\frac{t}{E}\right)\,,
\eea
is an effective parametrization of the tail of the high-energy peak, and $\bar{f}_{\mathrm{BG}}$ is given in Eq.~(\ref{eq:bgfitter}).
%

{The treatment of signal tails that we just described, especially when the tail is used to infer the peak position, to some extent makes our method more sensitive to the global shape of the energy spectrum. Therefore for the cases where tail contributions are important our method is less ``feature driven'' and more sensitive to the overall shape of the energy spectrum  and more exposed to issues related to our (mis)understanding of it. Despite the increased dependence on the overall shape of the energy spectrum, the information on the masses extracted with our method comes solely from the peak determination whereas the shape parameters, for instance $w_{LP}$ in eq.(\ref{templateLP}), are not used for the mass measurement, as, instead, one would do for a full-fledged shape analysis. Therefore we think that our analysis is quite distinct from a full-fledged shape analysis. With respect to such analysis, ours is still essentially based on the determination of a single feature of the distribution, a peak in our case, where the information is concentrated, as opposed to the  information diluted along {\it all} the distribution that a shape analysis would attempt to retrive.} 

\subsubsection{ Fitting of the energy spectrum for the mass spectrum II }
For the mass spectrum in which all three masses are comparable, we expect a $b$-jet energy spectrum similar to the one sketched in the right panel of Figure~\ref{toyspectra}. 
As can be seen therein, the two peaks in this case are largely overlapped and in fact the typical energy spectrum will have a single bump only. Armed with the knowledge of \eq{eq:fitter} we can extract the two component of the total energy spectrum and therefore measure the two peak locations even though the two peaks are not resolved. 
Since the two peaks are largely overlapping, we can concentrate our study on an energy range that includes only little part of the tails of the distribution.
Therefore for this type of spectrum there is no need for a special treatment of the tails, unlike for the energy spectra of the previous Section. Of course, we need to take into account the presence of background events in the data. Therefore, we take data points taken in a broad region around the the bump in the energy spectrum and we fit them with a function 
\bea
f_{\mathrm{HP}}(E)+f_{\mathrm{LP}}(E)+\bar{f}_{\mathrm{BG}}(E)\,,\label{fulltemplatecasetwo}
\eea
where the function $\bar{f}_{\mathrm{BG}}$ is defined above in Eq.~(\ref{eq:bgfitter}), $f_{\mathrm{HP}}$ is given in \eq{templateHP}, and $f_{\mathrm{LP}}$ is given in \eq{templateLP}.

{For this kind of spectrum we have to rely on our knowledge of the line-shape of each energy peak. Therefore we remark that this analysis goes beyond an energy peak analysis in the strictest sense. As a matter of fact the shape of the energy peaks plays an important role for our result and the reliability of the exponential functions $f_{\mathrm{HP}}$  given in \eq{templateHP}, and $f_{\mathrm{LP}}$  given in \eq{templateLP} is a key issue. In the following we show that these fit functions are good enough for our purpose, as demonstrated  by the results of the energy peak fit in eq.(\ref{e1e2}) in the next section.}

{For our single bump spectrum one might question how one can make sure that a this energy spectrum is originated by 2 two-body decays in each chain, as we will do, and not by another (simpler) process. In fact one could argue that a spectrum with a single bump could be originated by a single two-body decay   and that the super imposition of the spectra from signal and background can give rise to a shape perfectly matching the data. However the latter hypothesis can be discarded immediately looking not just at the overall energy distribution obtained looking at all the events, but also considering the characteristic of each event. In fact in our example each new physics event has 4 $b$ jets, therefore the most natural options in a $R$-parity conserving model is that the $b$ jets are either originated each in a two-body decay (as in our process) or by two decay chains each made of a single step three-body decay $\tilde{g}\to bb \chi$. The single two-body decay explanation just does not make sense on a event-by-event basis. Distinguishing wether the gluino decays in a chain of two body decays or in a single step three-body decays is more subtle. Most likely the two cases can be told apart looking at the $m_{bb}$ distribution, which should have a sharp edge for the cascade of two body decays and be much less sharp for a three body decay.} 

\subsection{Treatment of the dijet mass edge}
As discussed in Section~\ref{mainsection} to measure all the three masses we need to supplement the measurement of the two energy peaks with the measurement of a third observable. The dijet mass edge can be taken as an example of a third observable to close the system of equation and eventually obtain the masses as functions of the three observables. We remark that the dijet mass edge, as well-known~\cite{Hinchliffe:1999ve,Allanach:2000gf}, is influenced by combinatorial issues, which arise from the need to identify which pair of $b$-jets come from one gluino and what is the other pair of $b$-jets that comes from the other gluino. Several solutions to this problem have been proposed over the time~\cite{Albrow:1976jm,Hinchliffe:1997fk,Dutta:2011uq,Rajaraman:2011fj,Choi:2011lr,Curtin:2011ng}. Since the study of the dijet mass edge is not the central topic of our paper, we assume that these combinatorial issues can be addressed sufficiently well to not impact significantly on the results. This seems quite plausible for the process at hand. 
In fact, 
%
%
we checked that 
if one orders the $b$-jets by their transverse momentum and then constructs one dijet mass from the first and the fourth hardest and another dijet mass from the second and third hardest $b$-jet in the event, 
then about one half of the times the pairing is done correctly (this is true for both the example spectra). Furthermore,  the dijet mass spectrum obtained from the events where the pairing is done incorrectly is rather featureless. Therefore, we do not expect that the contribution from wrong dijet combinations will end up affecting the extraction of the edge of the distribution. 

In what follow we assume that the dijet mass edge can be extracted from the invariant mass distribution with high precision. Therefore, we neglect the propagation of the error on this measurement on the determination of the three masses of the new particles of our process \eq{gluinoTbbbb}. Our assumption about the error on the dijet mass may or may not be justified in specific  experimental situations. However, we prefer to not consider the error from the dijet mass edge in the error propagation  because in this way we put in full display the sources of error  that are characteristic of the novel analysis strategy that we propose in this paper.

\subsection{Results\label{Results}}

In this section we present our results about the mass measurement using the energy  peak fitting technique. We quantify the expected best-fit determination of the masses and the associated  expected uncertainty. To determine these quantities we take simulated samples of signal and background events corresponding to 300/fb of  $pp$ collisions at the 14 TeV LHC. The samples have been generated as described in section~\ref{simulation}. We take 100 samples corresponding to this luminosity, each being an iteration of a pseudo-experiment for the mass measurement. From each sample we derive the $b$-jet energy distribution after the cuts \eqs{preselection}-(\ref{dphicut}) and we fit the spectrum according to the fitting strategy described in Section~\ref{fittingstrategy}. From each experiment we determine the best-fit for the two peak energies $E_{LP}$ and $E_{HP}$ and we turn them into a mass measurement by mean of  \eqs{massesfromobservables} or some suitable approximation of them. From the same formulae we can propagate the fit uncertainties and obtain the error on the mass measurements. 
For our results we quote the expected mass measurement obtained by averaging the masses extracted in each of the pseudo-experiments. For the expected uncertainty on the masses we quote the average of the uncertainties obtained from each pseudo-experiment {looking at the $\chi^{2}$ variation of each fit}. 

\begin{figure}[h!]
\begin{center}
\includegraphics[width=0.49\linewidth]{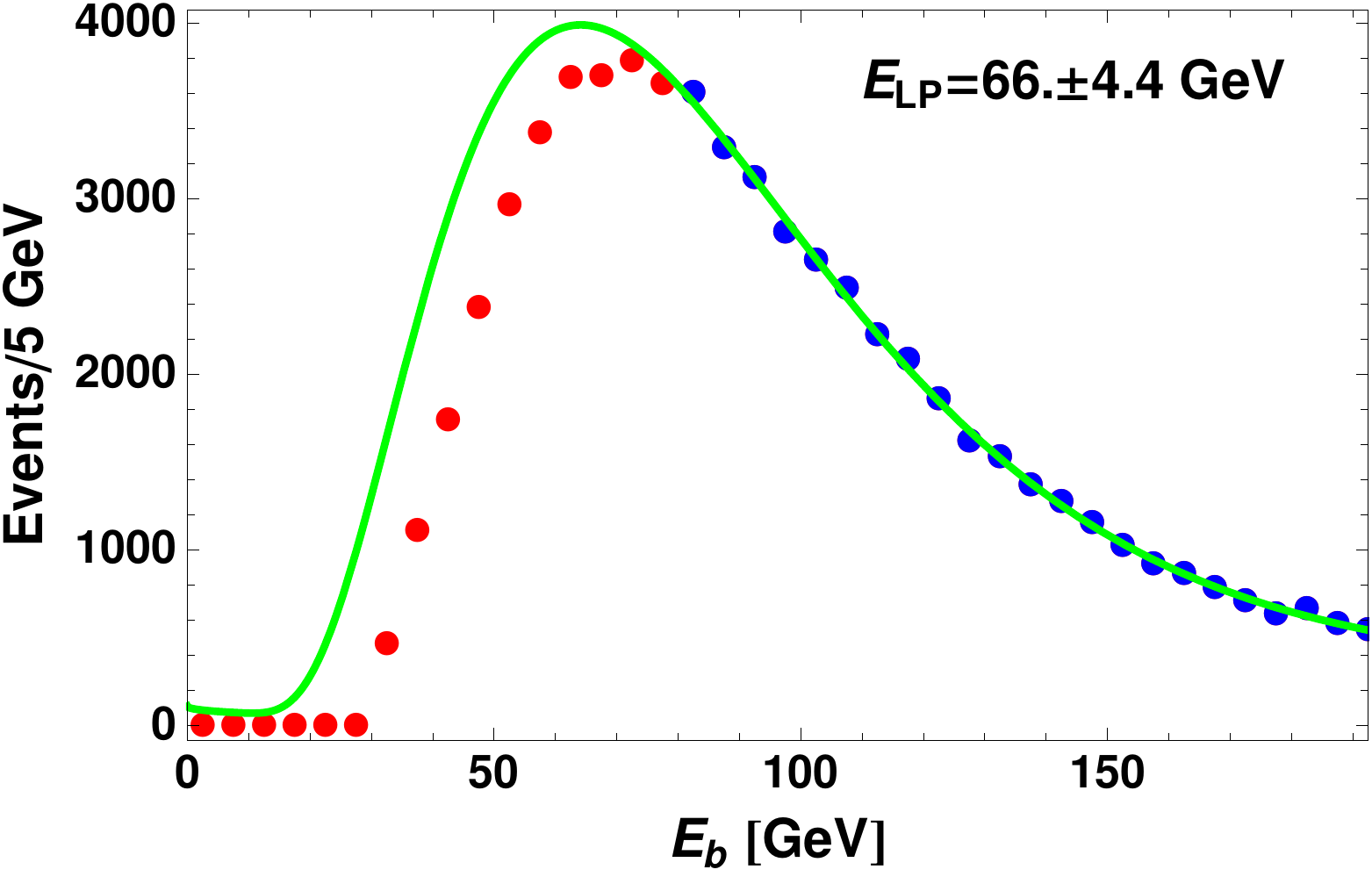}
\includegraphics[width=0.49\linewidth]{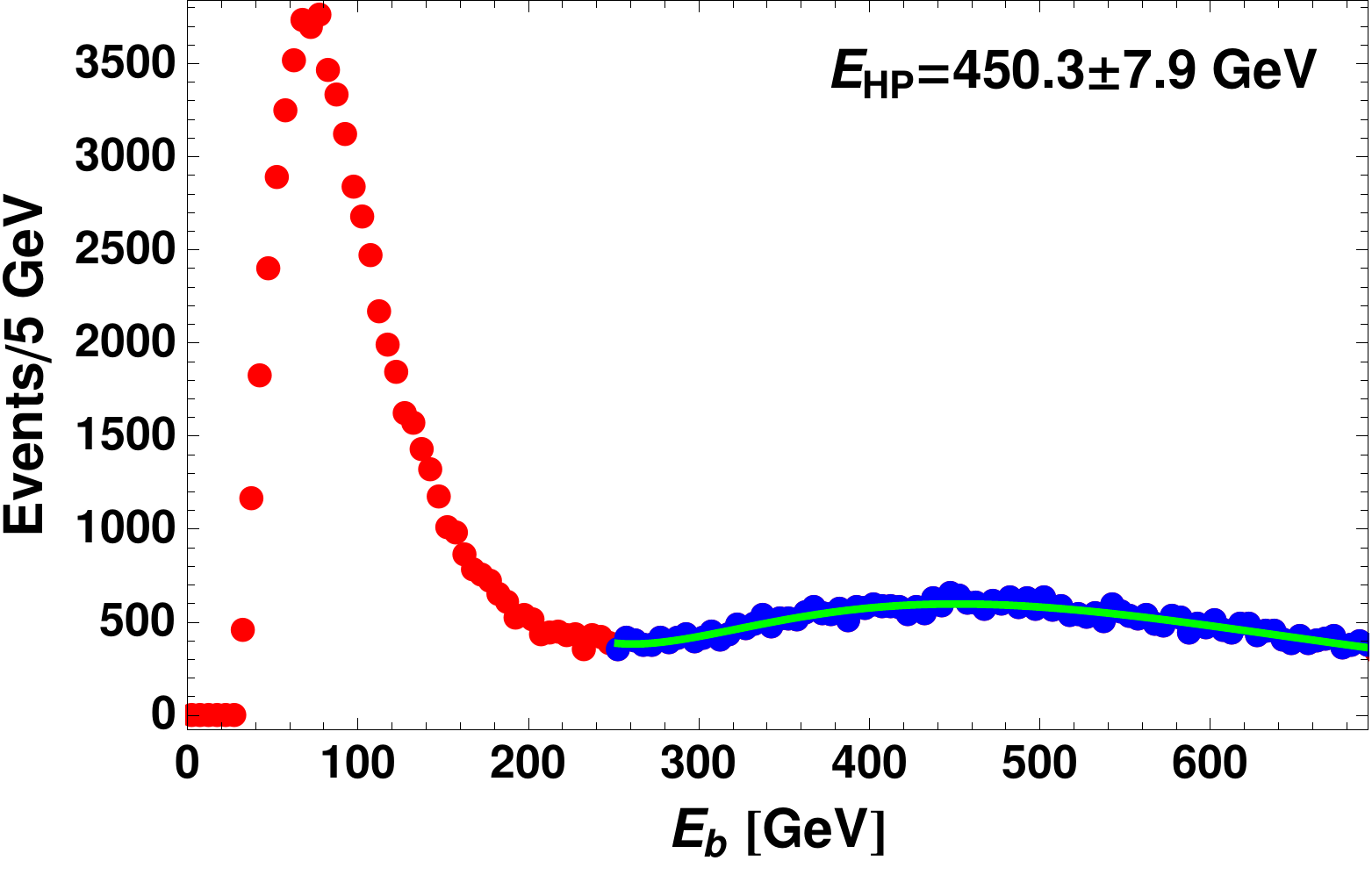}
\caption{Results of the energy peak fit on the data of representative pseudo-experiments for the spectrum I given by \eq{massesone}. The red and the blue dots are the data points after the cuts \eqs{preselection}-(\ref{dphicut}). The blue data points are those used to fit the function to the data. They correspond to the energy range 80-200 GeV, in the left panel, and 250-700 GeV in the right panel. The solid green line is the best-fit curve of the type \eq{LPtotaltemplate}, for the low energy range shown in the left panel, and \eq{HPtotaltemplate} for the high energy range shown in the right panel.}
\label{fig:repplots}
\end{center}
\end{figure}

\begin{figure}[h!]
\begin{center}
\includegraphics[width=0.49\linewidth]{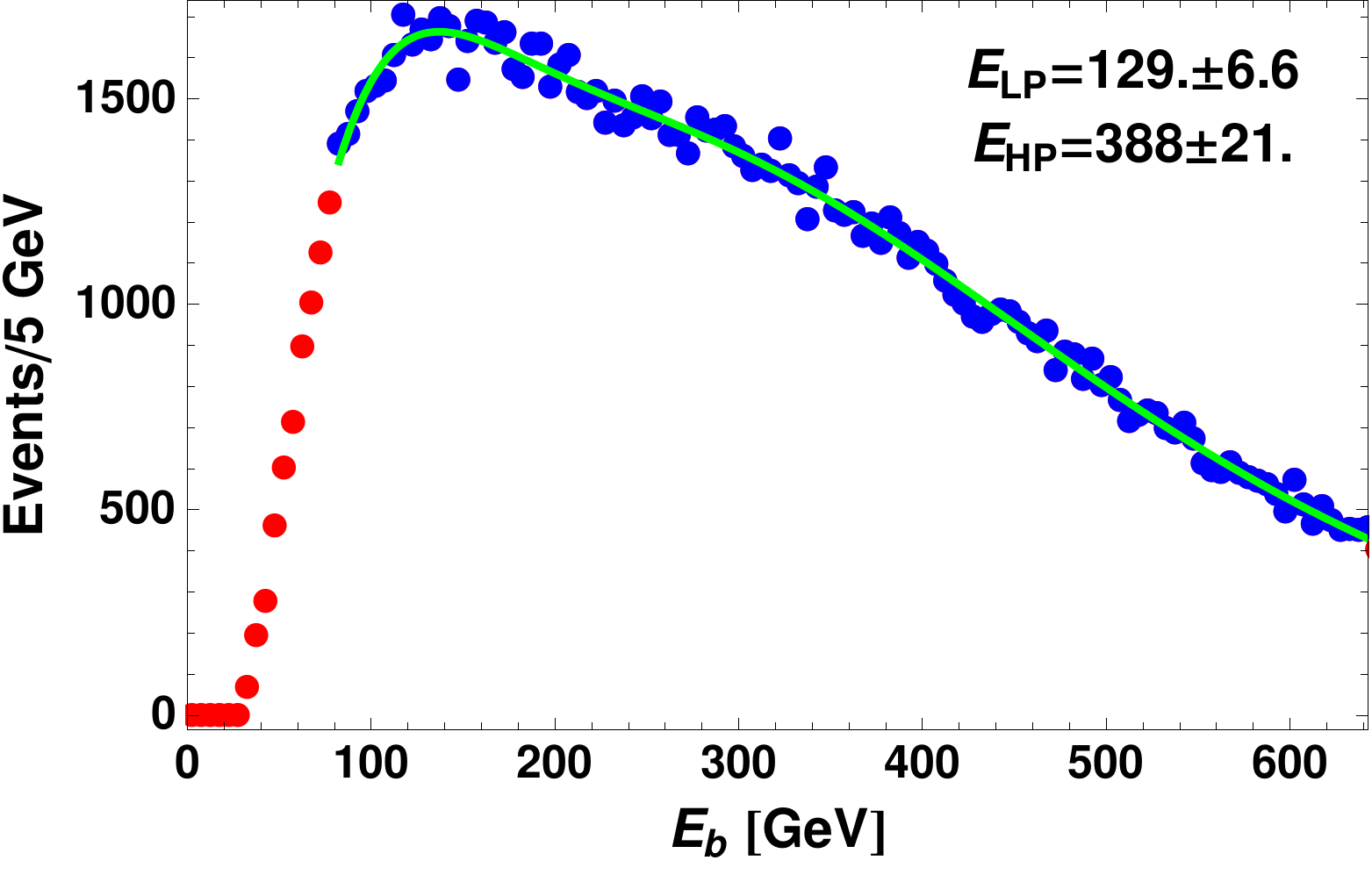}
\caption{Results of the energy peak fit on the data of representative pseudo-experiments for the spectrum II given by \eq{massestwo}. The red and the blue dots are the data points after the cuts \eqs{preselection}-(\ref{dphicut}). The blue data points are those used to fit the function to the data. They correspond to the energy range 80-650 GeV. The solid green line is the best-fit curve of the type \eq{fulltemplatecasetwo}.}
\label{fig:repplotstwo}
\end{center}
\end{figure}

Energy spectra and fitting results from a representative pseudo-experiment for the mass spectrum I defined in \eq{massesone} are shown in Figure~\ref{fig:repplots}. The darker data points (colored in blue) are those in the energy ranges actually used in the fit. As can be seen from the figure we fitted the two peaks of the spectrum I using the data points in the energy range 80-200~GeV for the low-energy peak and the energy range 250-700~GeV for the fitting of the high-energy peak. 
Similarly, in Figure~\ref{fig:repplotstwo} we show the fit results to the data from a representative pseudo-experiment for the mass spectrum II defined in \eq{massestwo}. In this case we used the data points in the energy range 80-650~GeV. {For both examples spectra we have tested that the choice of the energy ranges has not a significant impact on the resulting energy peak measurement, which is stable under variations of the chosen energy ranges for the fit.}

{Using 300/fb of $pp$ collisions at the 14~TeV LHC from the 100 pseudo-experiments for Spectrum I we obtain the following average measurement of the energy peaks
\beq
E_{LP}=68\pm7\gev,\quad E_{HP}=457\pm12\gev\,.
\eeq
For completeness, in Figure~\ref{onedimeoneetwo} we  report the one dimensional distributions of the measurement in each of the 100 pseudo experiments, from which one can get further information on the properties of the measurements of the energy peaks {\it per se}, i.e. not in connection with the interpretation of the energy peaks for the measurement of masses. 
We remark that the distribution of the pseudo-experiments for the low-energy peak fits has a variance that is quite smaller than the typical error from the $\chi^{2}$ profile analysis in the fit. 
Both assessments of the error on the measurement are below 10\% level and for the scope of this exploratory paper, where we do not pursue a high precision measurement, 
we do not investigate the meaning of the deviation of the distribution of the pseudo-experiment fit results from that of the $\chi^{2}$ analysis.

\begin{figure}[t!]
\begin{center}
\includegraphics[width=0.49\linewidth]{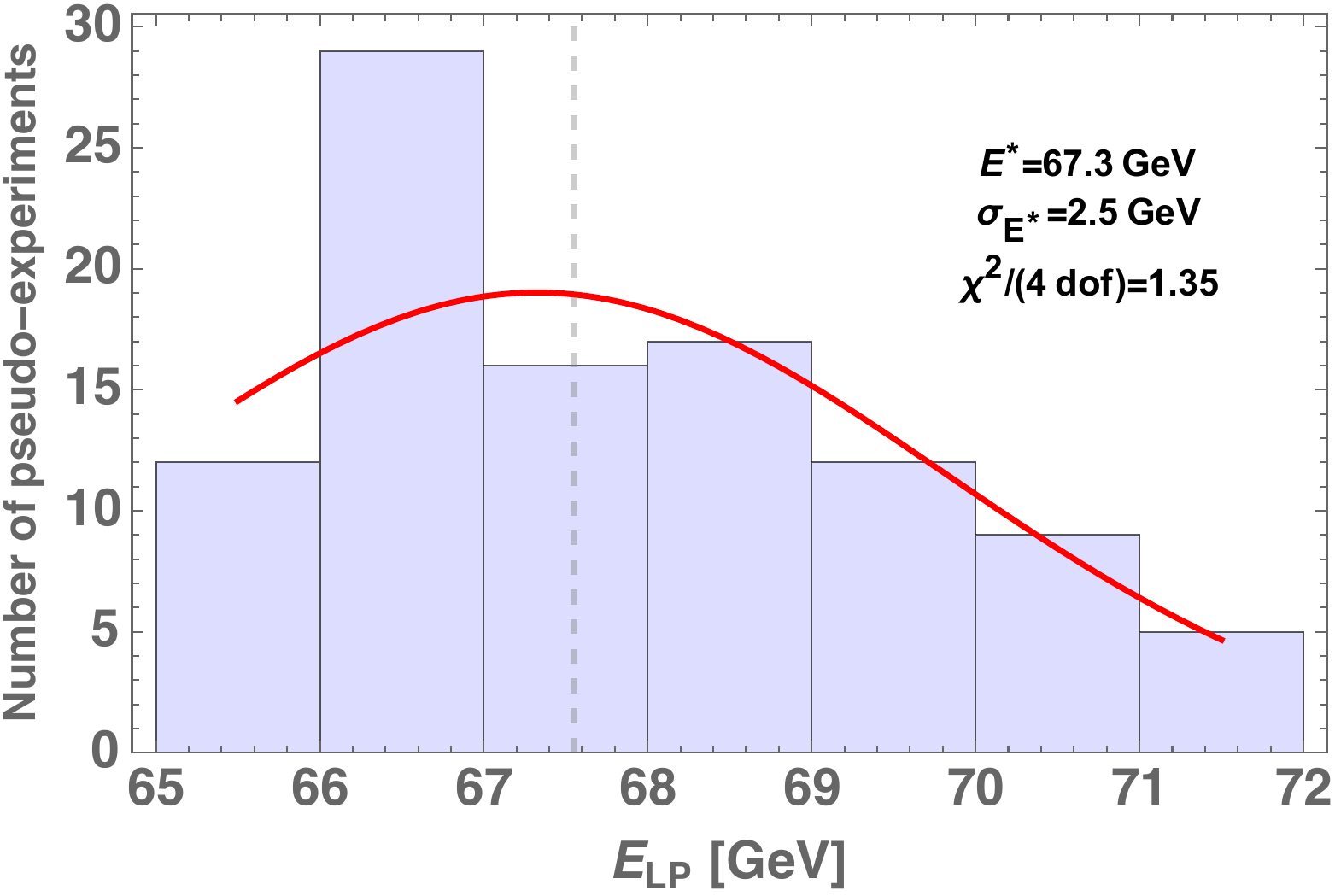}
\includegraphics[width=0.49\linewidth]{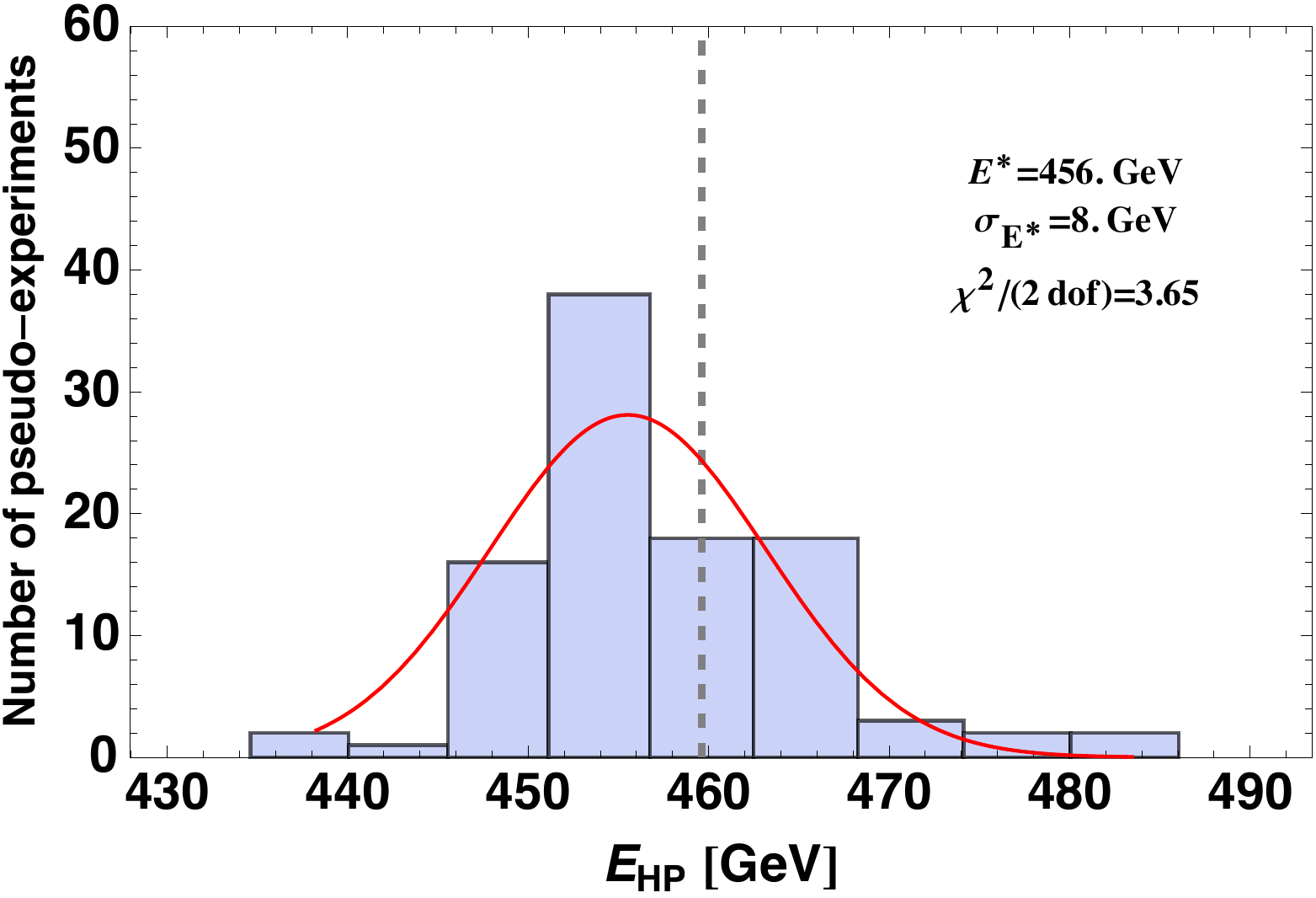}

\caption{Two panels for the  one-dimensional distribution of the $E_{LP}$ and $E_{HP}$ obtained from the fits of the two separate energy peaks over 100 pseudo-experiment for the spectrum given by \eq{massesone}. The vertical dashed lines represent the true value from theory. We also report the parameters of a fit to the distribution of the pseudo-experiments results with a gaussian, that we use as a check of the normality and bias of the energy peak fitting procedure.}
\label{onedimeoneetwo}
\end{center}
\end{figure}

{Using the energy peaks measured in our fits in the 100 pseudo-experiments and the relations \eq{massesfromobservables} we obtain an not very meaningful average mass measurement. In fact the uncertainties on the masses  are of order TeV, hence too large for being of any interest. In order to extract more accurately some of the masses for the spectrum I we can proceed as outlined in Section~\ref{spectrumone} and we begin by taking a massless neutralino.}

Fixing $m_{\chi}=0$ we can choose which pair of observables to be used to obtain the masses of the gluino and the sbottom. That is to say, we can either use the two peak energies $E_{LP}$ and $E_{HP}$ extracted from the fit and obtain the masses from the relations \eq{simplifiedcaseoneenergy}, or alternatively we can choose to use the high-energy peak $E_{HP}$ from the fit together with the dijet mass edge and obtain the masses from  \eq{simplifiedcaseone}.

From the same 100 pseudo-experiments for Spectrum I we expect a mass measurement obtained from the two energy peaks that is 
\beq
m_{\tilde{g}}=986\pm63\GeV,\quad m_{\tilde{b}}=919\pm37\gev\,,  \label{expectedmasscaseoneedge}
\eeq
whereas using the high energy peak and the dijet mass edge we expect
\beq
m_{\tilde{g}}=989\pm34\GeV,\quad m_{\tilde{b}}=919\pm37\gev\,.  \label{expectedmasscaseoneE}
\eeq
The use of the dijet mass edge clearly improves the mass measurement for the gluino, while it has no impact on the sbottom mass determination. This is simply explained by comparing \eq{simplifiedcaseone} and \eq{simplifiedcaseoneenergy} for the relation between the observables and the masses. Additionally we remark that, considering the associated nominal values, i.e., 1000~GeV and 930~GeV, the measured values are in quite a good agreement within $1\sigma$ range for both approaches.

{For the neutralino mass determination we can exploit the gained knowledge on $m_{\tg}$ and $m_{\tb}$ from which we have derived \eq{mchiexactfunctionofpeaks}. This equation can be used to determine the neutralino mass in each pseudo-experiment  and the average neutralino mass measurement in this case is 
\beq
m_{\chi}=\textrm{sign}(m_{\chi}^{2}	)\sqrt{\left| m_{\chi}^{2}\right| }=117\pm 366 \gev\,.
\eeq
The expected uncertainty on the mass measurement agrees with  the  expectation from the analysis summarized in Figure~\ref{mChivsElpEhp} and allows us to disfavor neutralino mass larger than about 500 GeV.}

\begin{figure}[t!]
\begin{center}
\includegraphics[width=0.49\linewidth]{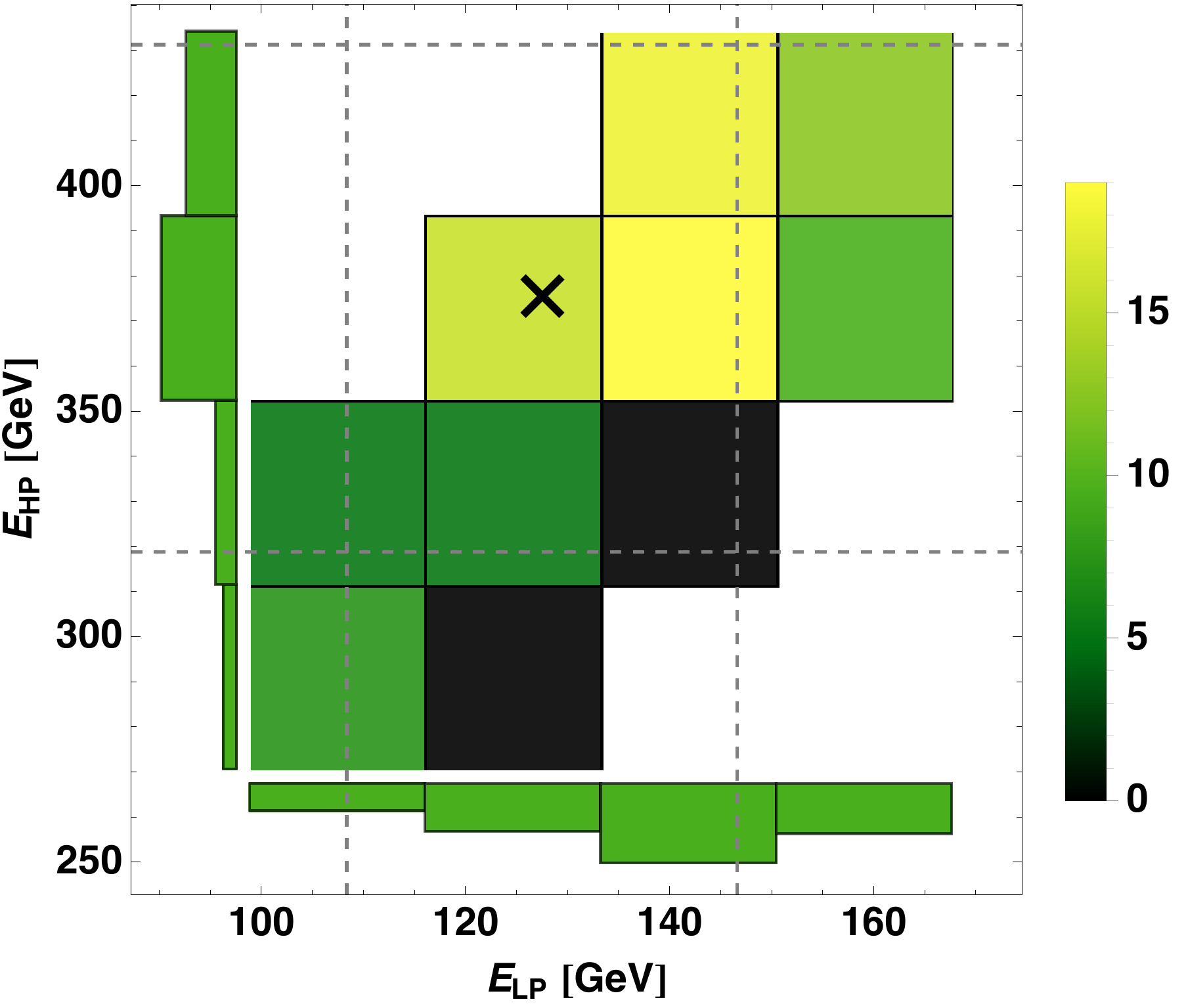}
\caption{Distribution in the plane $E_{LP},\,E_{HP}$ for the fit of the two energy peaks over 100 pseudo-experiment for the spectrum given by \eq{massestwo}. The black cross represent the true value from theory and the dashed lines delimit the region of the plane within a 15\% variation from the theory value. On the vertical (horizontal) axis we report the unidimensional distribution of $E_{HP}$ ($E_{LP}$).}
\label{eoneetwo}
\end{center}
\end{figure}

For the spectrum II the three masses are comparable, as reflected by the fact that the two energy spectra from the two steps of the decay are largely overlapped and result in a energy spectrum with a single bump. 
{Using 300/fb of $pp$ collisions at the 14~TeV LHC the expected energy peaks measurement for Spectrum II is
\beq
E_{LP}=137\pm 17 \gev,\quad E_{HP}=372\pm 36\gev\,.
\label{e1e2}
\eeq
For completeness in Figure~\ref{eoneetwo} we also report the one- and two-dimensional distributions of the energy peaks measurement  in 100 pseudo experiments. These distributions are useful to get further information on the properties of the measurements of the energy peaks {\it per se}, i.e. not necessarily in connection to the interpretation of the energy peaks for the measurement of masses. From the figure we observe that the distribution of $E_{LP}$ and $E_{HP}$ tends to have a moderate correlation.}

Given that none of the masses can be neglected, in order to extract them  we use  the exact relations \eqs{massesfromobservables}  taking as  input observables the dijet mass edge and the peak energies $E_{LP}$ and $E_{HP}$ from the fit.
Turning the energy peaks measurement in mass measurement we obtain the expected measurement 
\beq
m_{\tilde{g}}=935\pm258\GeV,\quad m_{\tilde{b}}=439\pm343\gev\,,\quad m_{\chi}=446\pm464\gev\,. \label{expectedmasscasetwo}
\eeq
As we can see the errors on the masses are quite large. {This might be understood from the results shown in Figure~\ref{contraintsplanecasetwo}. In the figure we can appreciate how for $m_{\chi}$ a variation of 100~GeV around the true value is not nearly enough to get the constraint to outside the 10\% error band, especially because the constraint (in blue) from the gluino decay is satisfied on a pretty wide band of the plane.} Ultimately we can ascribe the size of the errors  to the high powers of the peak energies that enter in the
%
%
expressions for the masses in terms of the observables \eqs{massesfromobservables}. 
In turn, the solutions for the masses in \eqs{massesfromobservables} are inversions of 
highly non-linear relations between the masses and the quantities 
$m_{bb}$, $E_{LP}$, and $E_{HP}$ given in \eqs{eq:peaka}-(\ref{eq:invedge}). 
Therefore, we believe that such large uncertainties on the masses are somewhat expected. 
Furthermore, we remark that the non-linearity of the relations between the masses and the observables is to some extent unavoidable unless some constraint or approximation can be used to simplify these relations~\footnote{In a way, this observation justifies the smaller errors obtained in \eq{expectedmasscaseoneedge}, where the neutralino is assumed to be massless. 
In fact, the relevant simplified equations  \eq{simplifiedcaseone} used to obtain such a result are far closer to be  linear  than the full system of  \eqs{massesfromobservables}.  }. 
The reason is that the quantities $E_{LP}$ and $E_{HP}$ are the only dimensionful quantities fixed by four-momentum conservation in each two-body decay. 
Therefore, we expect that such large uncertainties will appear in the results of typical mass measurement techniques developed for such multiple two-body decay chains.

In this respect, we remark that most of the non-linearity of \eqs{massesfromobservables} arises from the dijet mass edge relation. Therefore, we find it instructive to consider what would be the expected mass measurement in case we can avoid to use the dijet mass edge. In order to do that, we need to fix the correct mass of one of the new physics particles. We choose to fix the neutralino mass in the system of equations. This choice is motivated by the fact that, unlike the other colored particles that we consider, the mass of the neutralino could be in principle measured along with its discovery in other experiments such as direct or indirect Dark Matter detection experiments.
Fixing $m_{\chi_{1}^{0}}=350\gev$ we can reduce  \eqs{massesfromobservables} to a 2-by-2 system of equations that can be solved for the gluino and the sbottom masses. In this case we decouple the dijet mass edge equation such that we can measure the two masses from the two energy peaks. The expected mass measurement is
\beq
m_{\tilde{g}}=978\pm70\GeV,\quad m_{\tilde{b}}=495\pm19\gev\,, \label{expectedmasscasetwomchi}
\eeq
which is substantially more accurate that what we obtain in case all the three masses have to be obtained from the fit. Again, we see that they are consistent with their corresponding nominal values, i.e., 1000~GeV and 500~GeV, within $1\sigma$ range.

Before concluding this section we comment on possible variations of $S/B$. In the SUSY example that we have considered the signal rate after the event selections is much larger than that of the backgrounds. This situation is particularly favorable for the mass measurement. However, we would like to test the robustness and the accuracy of our energy peak fitting mass measurement in a more general context than just the concrete gluino decay example that we have discussed. To this end we have repeated the mass measurement on several sets of 100 pseudo-experiment. In each group of 100 pseudo-experiment we have deliberately increased the cross-section of the background by some factor, such as to get a lower $S/B$ and therefore worse conditions for the mass measurement.
We have checked that, despite the less favorable $S/B$, our results are stable under modest changes of the background cross-section. Furthermore, we have investigated how the mass measurement degrades when the background cross-section is enhanced by a large factor. We observe that the result slowly degrades as $S/B$ gets smaller and eventually the fit errors become comparable with measured peak energies once $S\simeq B$ is reached.

\section{Outlook and conclusions\label{conclusion}}
The Standard Model has been a successful description for fundamental interactions in Nature. Nevertheless, several questions such as the Planck-Weak hierarchy and the Dark Matter problem still remain unanswered. Extensions of the SM invoked to solve these problem typically involve new particles at the TeV scale. In this context, the ongoing experiments at the Large Hadron Collider  are expected to discover new particles in the near future. Once such discoveries are made, one of the natural questions to ask next is to determine the physics parameters of such new particles such as their coupling constants, gauge charges, spin, mass etc.    

In this paper we studied the mass measurement of the new physics particles that are involved in a two-step cascade decay chain which includes a (massive) invisible particle along with other SM final states (see \eq{genericdecay}). In these decay chains there are in principle three relevant new particles masses. In order to determine them we use two independent relations from the energy distributions of visible particles. To the best of our knowledge this is the first time that the energy distribution is used in this way to get information about the masses involved in the decay. 
This relies crucially on the observation that the location of the peak in the energy distribution of the visible particles coming from a given two-body decay is actually the same as its energy measured in the rest frame of its immediate mother particle~\cite{Agashe:2012mc,1971NASSP.249.....S}.
To have a chance to measure all the three unknown masses we supplement the relations coming from the energy distribution with a third independent relation from the well-known edge in the invariant 
mass distribution of the two visible particles. 

As a concrete example  for demonstrating the {\em general} mass measurement technique proposed in this paper, we studied simulated LHC events from the production of pairs of gluino, followed by its decay of  into two $b$ quarks and the lightest neutralino via an on-shell bottom squark (as shown in \eq{gluinoTbbbb}), in the context of SUSY with conserved $R$-parity.
Since the neutralino is {\it invisible} (and massive),
the final state signature on which we concentrate is $4b+\misse$.  
The visible particles of this signature are the $b$ quarks that are emitted from the decay of the gluino and the sbottom. Since in the final state there are four $b$ quarks it is hard to distinguish which $b$ quarks in each event are originated at each of the steps of the decay chains. Therefore, in order to retrieve the energy peaks associated with each step of the decay chain we are forced to study the {\it inclusive} energy distribution of the $b$-jets, to which, for each event,  the energies of all the four $b$-jets contribute.
For more systematic analyses we investigated in detail two representative scenarios of mass spectrum. These are characterized by  the different relative distance between the two peaks in the energy distribution as shown in the schematic decomposition of the energy spectra of Figure~\ref{toyspectra}. In case of mass spectra that have degenerate masses, we expect to see well-separated peaks. On the other hand, for mass spectra that have all the masses of comparable size, we expect a single bump in the energy spectrum. 

The core of our mass measurement strategy is the determination of the location of the peaks  by fitting the simulated experimental data with a fitting function suitably designed to address the features of each type of energy spectrum. The basic function used to identify the peaks is the same as that proposed in our earlier work~\cite{Agashe:2012mc}. 

To deal with the backgrounds originating from the SM processes, we imposed  the cuts \eqs{preselection}-(\ref{dphicut}). We found that the process $pp\to Z+4b$, where $Z$ subsequently decays into two neutrinos, is the dominant background for our  signature. In our fit to the simulated energy spectrum we have introduced a suitable function to take into account the presence of events from the background processes.

For the energy spectrum where the two peaks are well-separated, which we denoted as Spectrum I given in \eq{massesone}, we extracted the two peaks by doing two separated fits in two different energy ranges chosen to isolate one peak at a time. In each fit we take into account simultaneously the peak that dominates in that energy range, the background from the SM processes, and the contamination effects from the tail that comes from the other peak present in the signal. For this type of energy spectrum we showed that even using the events that come from the higher-energy tail of the lower  peak can be sufficient to reconstruct the position of the peak. This is possible thanks to the reliability of the peak template \eq{eq:fitter} that we have introduced in~\cite{Agashe:2012mc}. This possibility proves particularly useful when the data around the peak region is either cut away or strongly biased by the event selections that are necessary to isolate the new physics events from the backgrounds.  
Furthermore, studying Spectrum II given in \eq{massestwo}, we have also shown that the  peak template \eq{eq:fitter}  allows to extract the values of the two peaks from an energy spectrum that clearly shows a single bumpy feature. This spectrum arises from two largely overlapping energy spectra from each step of the decay chain and we showed that with sufficient statistics the position of the two distinct peaks can be disentangled.

The final results for the expected mass measurements under different assumptions for the spectrum and different treatment of the observables to get the mass measurement are collected in \eqs{expectedmasscaseoneedge}-(\ref{expectedmasscasetwomchi}).
Overall we see that the masses can be determined with a relative precision up to about few times 10\%, which seems a rather encouraging sign that the novel mass measurement technique presented in this paper is useful. 

Beyond the example that we consider in detail our mass measurement technique can be used on a variety of new physics processes. In fact, the fundamentals of our strategy is our result on the position of the peak in the energy distribution of  massless  decay products in a two-body decay and the effective description of its shape around the peak~\cite{Agashe:2012mc}. 
In particular, mass measurement strategies similar to the one described in our paper can be conceived for particles that, at variance with the example discussed here, are singly produced. Furthermore, the technique can be easily generalized to longer decay chains. 

More generally speaking we would like to remark that the ideas exposed in this paper can find applications beyond the problem of mass measurement. For instance in \cite{Low:2013kx} it was observed that the solid expectations on the 
location of the peak 
of the energy spectra in two-body decays that stem from our result~\cite{Agashe:2012mc} can be used to better isolate signals of new physics from SM background processes. Therefore, we believe that our observation has some potential to improve both current searches for new physics and the measurement of the masses of the new particles to be discovered at the LHC. Keeping {\em both} these two goals in mind, we envisage and look forward to further novel applications of our results, including applications in conjunction with other techniques~\cite{Barr:2011ao,Rogan:2010kb,Kawabata:2011gz,Kawabata:2013fta,Kim:2010lr}. 

\section*{Acknowledgements}
We would like to thank Matthew Buckley, Roberto Contino, Sarah Eno, Nicholas Hadley, Joseph Lykken, Konstantin Matchev, Michele Papucci, Tuhin Roy, Raman Sundrum, Jeff Temple, Jesse Thaler, Lian-Tao Wang and Kyle Wardlow for discussions and
Hongsuk Kang, Myeonghun Park, Jeff Temple, and Young Soo Yoon for help with fitting.
This work was supported in part by NSF Grant No.~PHY-0968854.
%
%
The work of R.~F. is also supported by the NSF Grant No.~PHY-0910467, and by the Maryland Center for Fundamental Physics.
R.~F. acknowledges the hospitality of the Aspen Center for Physics,
%
%
which is supported by the 
NSF
%
%
Grant No. PHYS-1066293. 
R.~F. also thanks the Galileo Galilei Institute for Theoretical Physics for the hospitality and CERN Theory Division and INFN for partial support during the completion of this work.
The work of D.~K. is also supported by DOE Grant No.~DE-FG02-97ER41029 and
the LHC Theory Initiative graduate fellowship (NSF Grant No.~PHY-0969510).

\appendix
\section{Variations of the background rate}

The study discussed in the main text takes background normalizations from actual predictions of the Standard Model for the signal and background rates.
For completeness of illustration of our technique, we show also fit results when the relevant background is not normalized to the prediction of the Standard Model. We perform similar analyses with inflated background rates that are 10 and 100 times larger than the prediction of the Standard Model. These analyses serve the purpose of assessing the performance of our technique when the signal to background ratio is less favorable than the $S/B\sim100$ that applies for the study in the main text.

The results are demonstrated in Figures~\ref{case1e1bg10and100} (Spectrum I) and~\ref{eoneetwobg10and100} (Spectrum II). Comparing the results for different $S/B$ shown in these figures, and also comparing them with the main results in Figures~\ref{onedimeoneetwo} and~\ref{eoneetwo}, one can see that there is no significant degradation of the fit results in the range of $S/B$ from 1 to 100 probed in these analyses.

\begin{figure}[ht!]
\begin{center}
\includegraphics[width=0.49\linewidth]{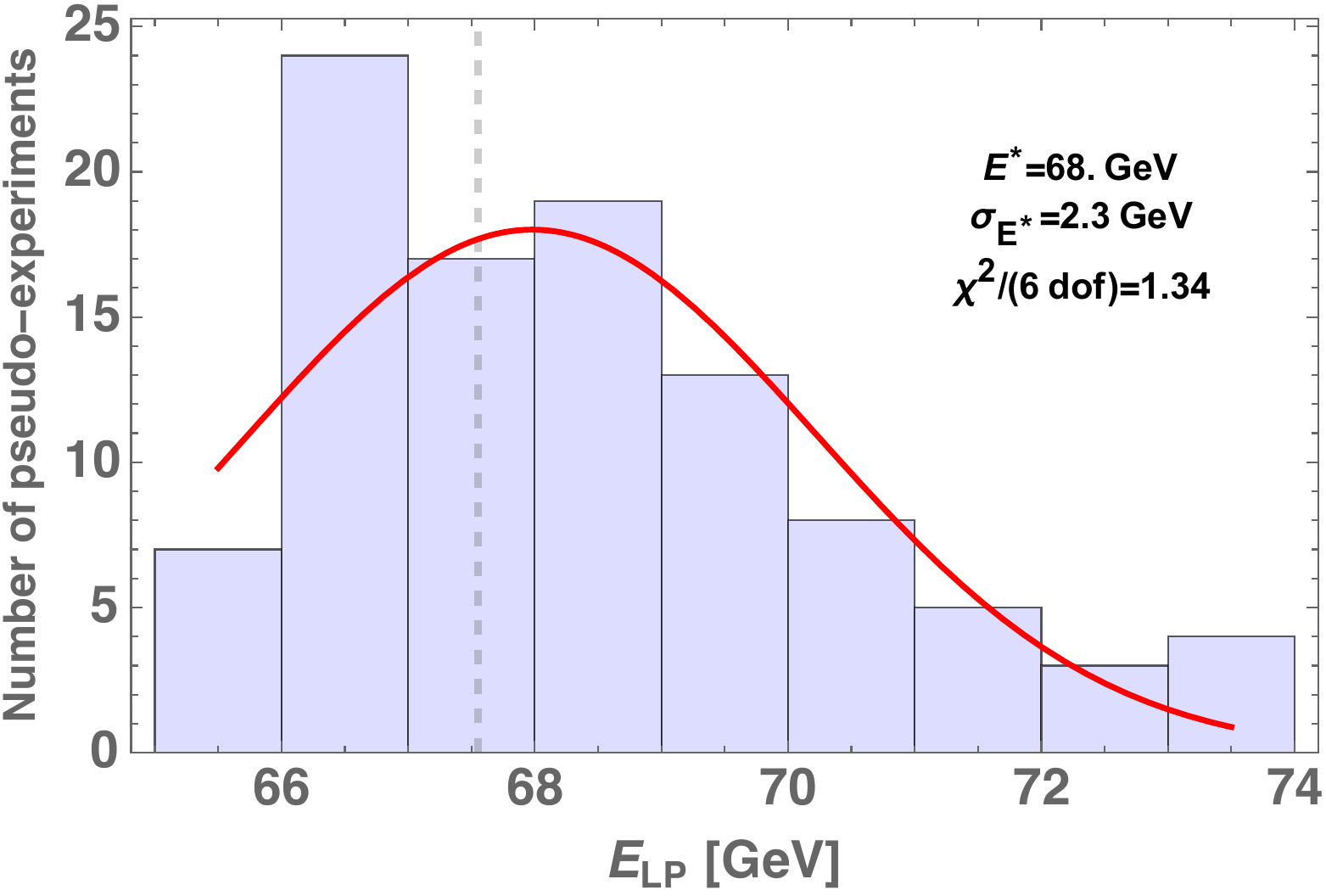}
\includegraphics[width=0.49\linewidth]{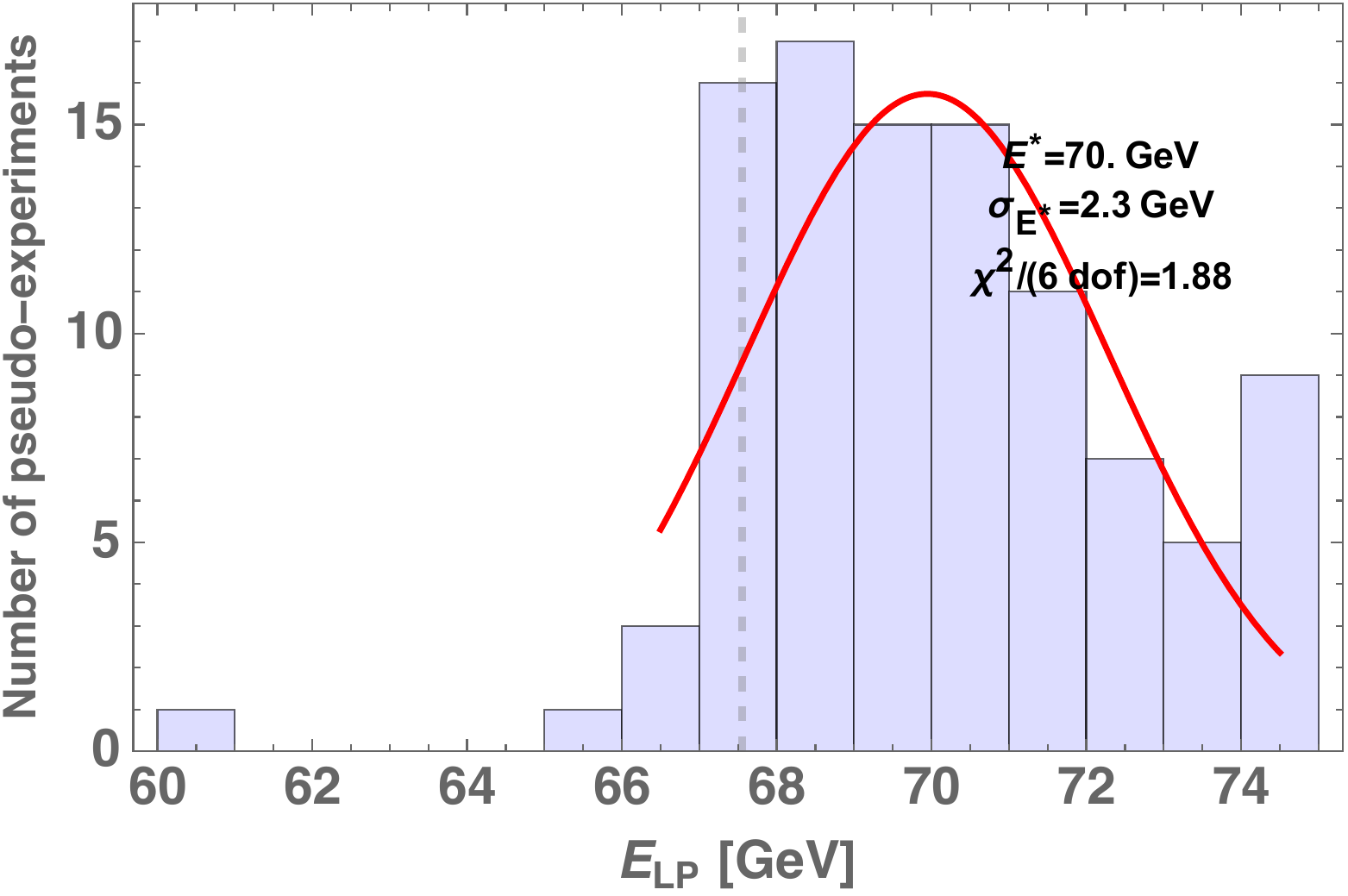}\\
\includegraphics[width=0.49\linewidth]{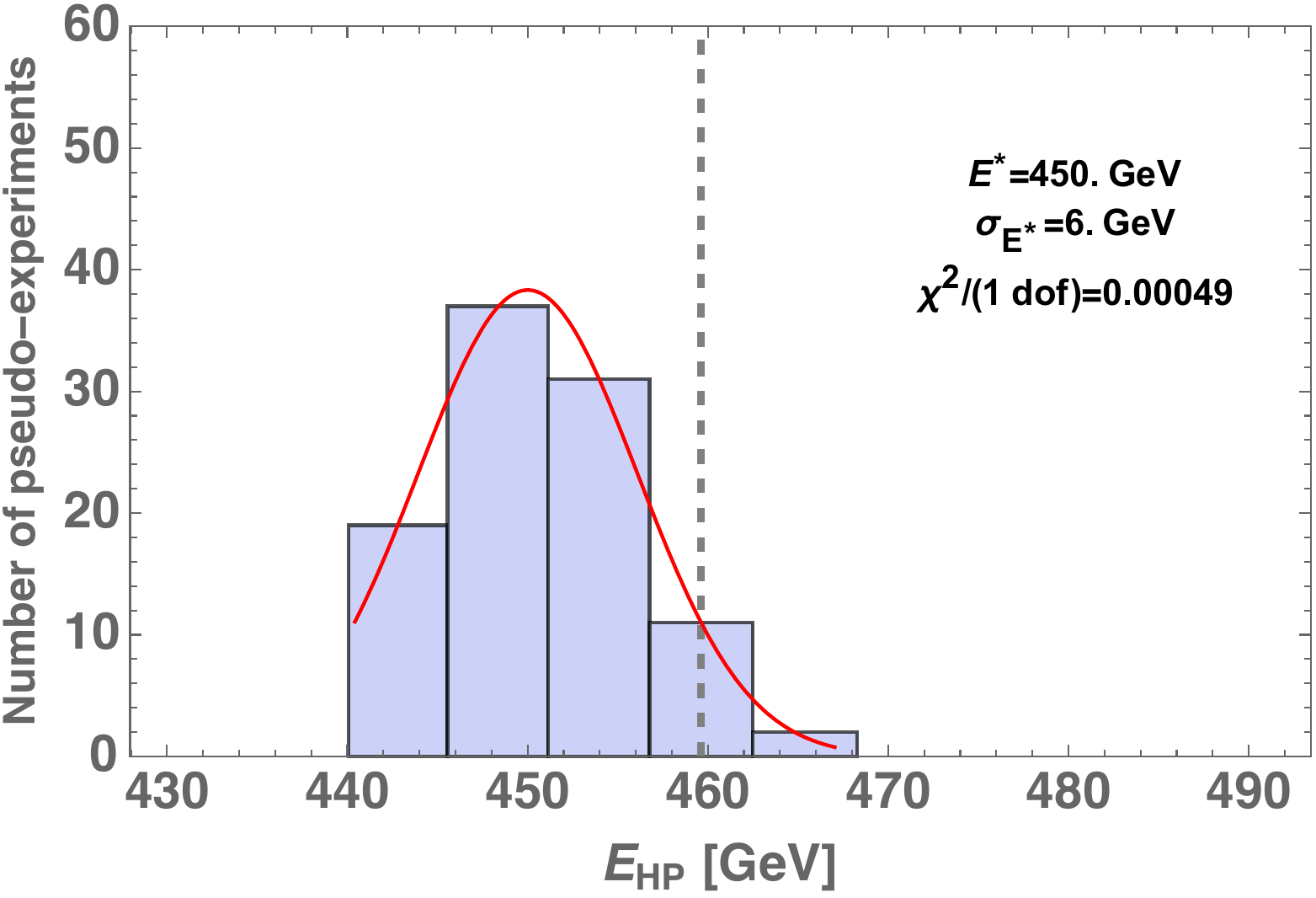}
\includegraphics[width=0.49\linewidth]{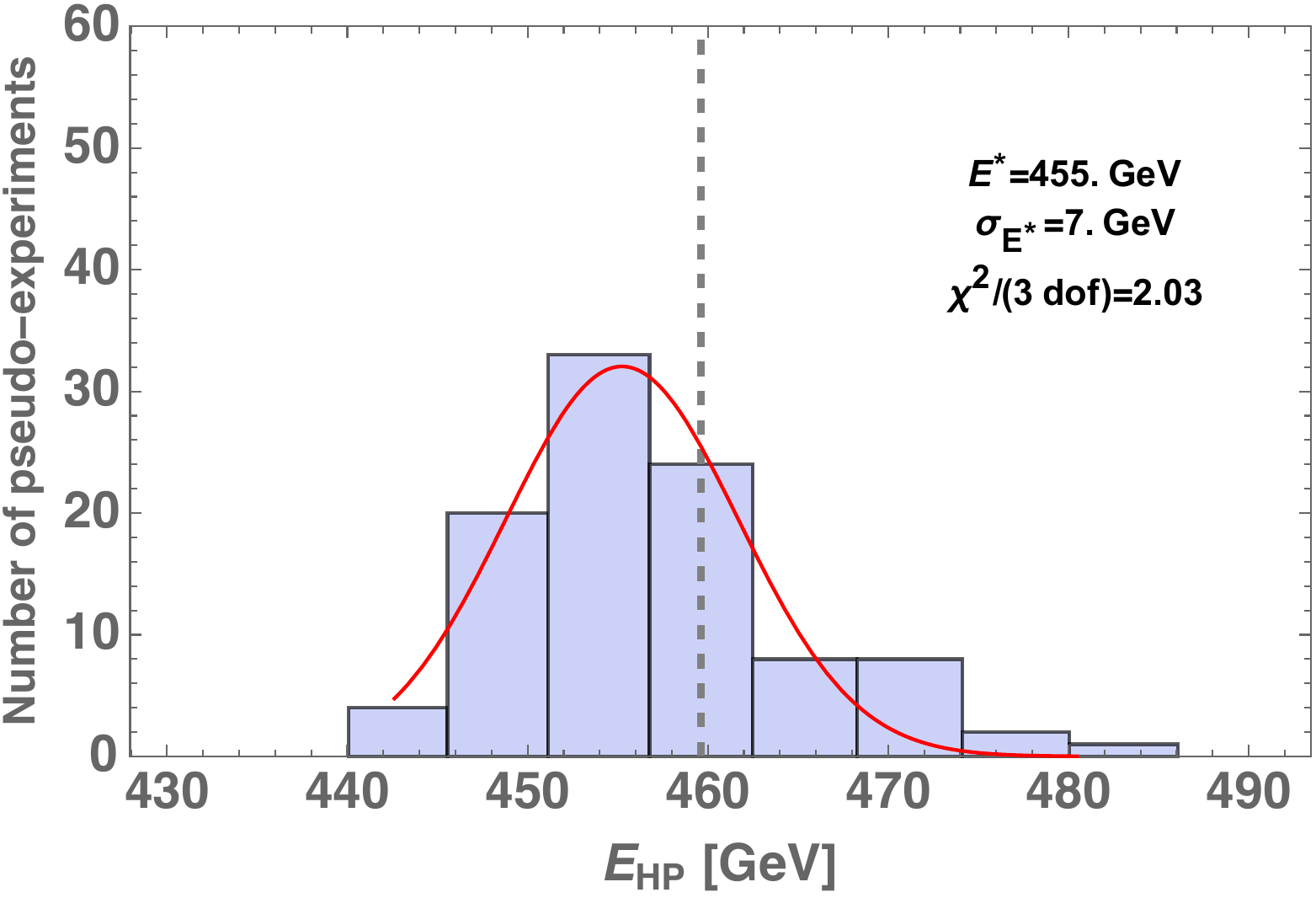}

\caption{Four panels for the one-dimensional distribution of the $E_{LP}$ and $E_{HP}$ obtained from the fits of the two separate energy peaks over 100 pseudo-experiment for the spectrum given by \eq{massesone}. The top line is for $E_{LP}$, the bottom line is for $E_{HP}$. Left and right columns are for fits with the background inflated by a factor 10 and 100, respectively, compared to the actual Standard Model rate. The inflated background rates correspond to a decreased $S/B\sim10$ and $1$, respectively. The vertical dashed lines represent the true value from theory. We also report the parameters of a fit to the distribution of the pseudo-experiments results with a gaussian, that we use as a check of the normality and bias of the energy peak fitting procedure.}
\label{case1e1bg10and100}
\end{center}
\end{figure}

\begin{figure}[t!]
\begin{center}
\includegraphics[width=0.49\linewidth]{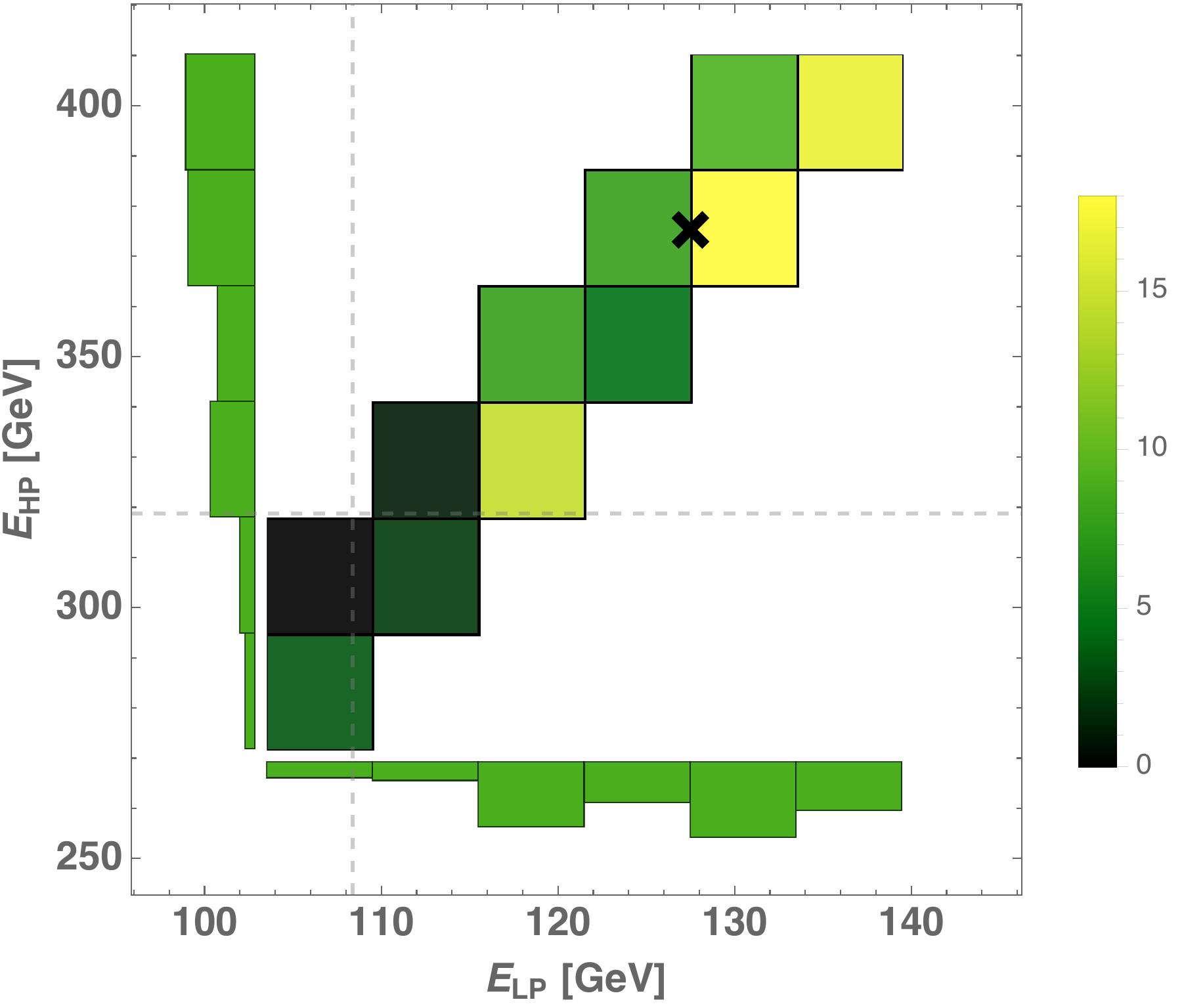}
\includegraphics[width=0.49\linewidth]{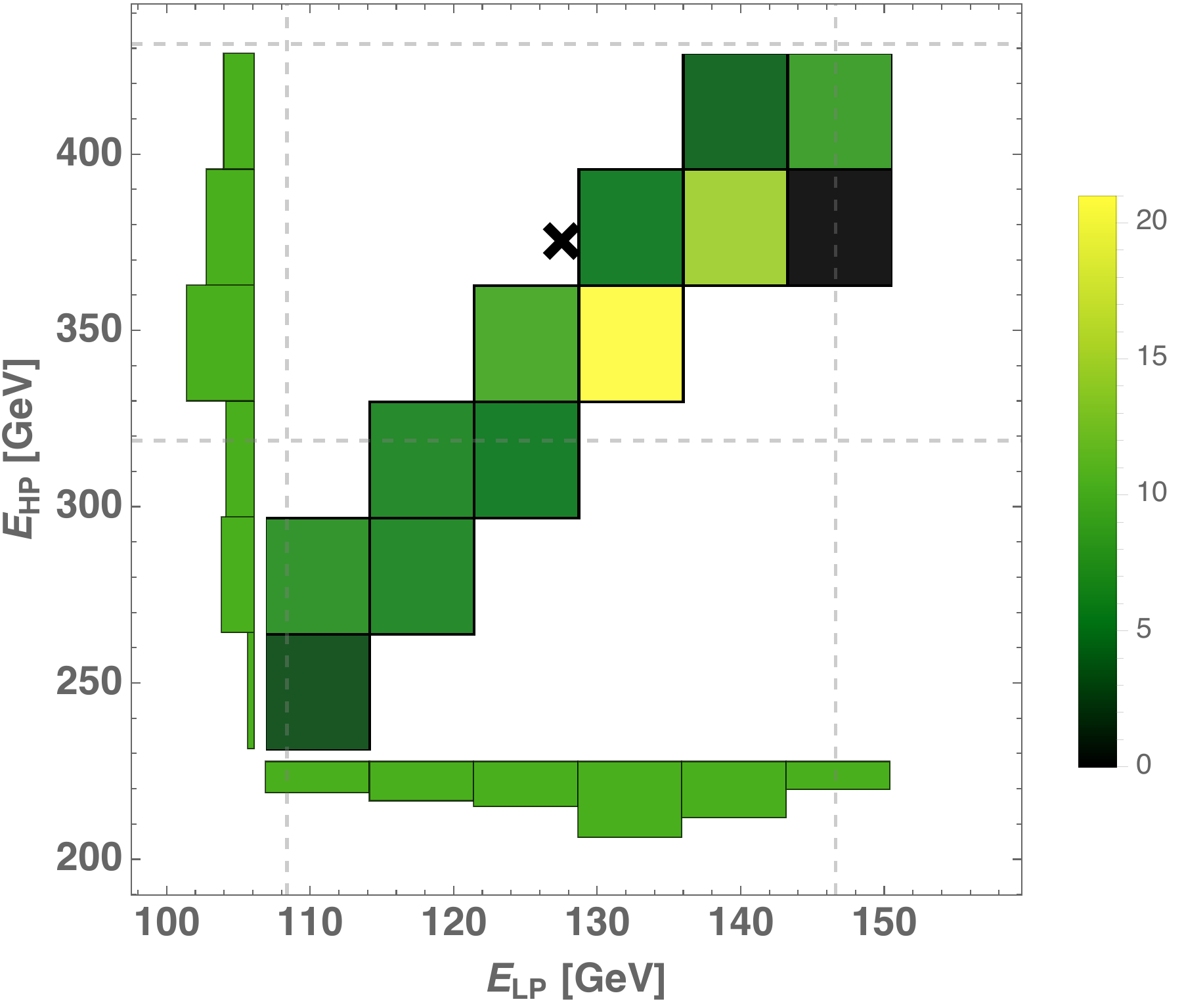}
\caption{Distribution in the plane $E_{LP},\,E_{HP}$ for the fit of the two energy peaks over 100 pseudo-experiment for the spectrum given by \eq{massestwo}.  Left and right panels are for fits with the background inflated by a factor 10 and 100, respectively, compared to the actual Standard Model rate. The inflated background rates correspond to a decreased $S/B\sim10$ and $1$, respectively.  The black cross represent the true value from theory and the dashed lines delimit the region of the plane within a 15\% variation from the theory value. On the vertical (horizontal) axis we report the unidimensional distribution of $E_{HP}$ ($E_{LP}$).}
\label{eoneetwobg10and100}
\end{center}
\end{figure}

Increasing the background normalization such that $B\gg S$, we expect a breakdown of the applicability of our method. However, we do not further investigate the exact value of $S/B$ where such breakdown happens. The reason is the following. When the background is much larger than the signal it is the understanding of the background that mostly determines the quality and the uncertainty of the measurement. The modeling of the background is not universal, and thus needs to be studied on a case-by-case basis for each specific mass measurement. Given the several {\it case-specific} complications that the analysis for the case  $B\gtrsim S$  would bring in our analysis, we think that such background dominated scenarios are not very informative about the technique presented here.


\end{document}